\documentclass{revtex4-1}
\usepackage[english]{babel}
\usepackage{amsmath}  
\usepackage{amsfonts} 
\usepackage{graphicx}

\begin{document}
\selectlanguage{english}

 \title{Time evolution of two-dimensional quadratic Hamiltonians: A Lie
algebraic approach.}


\author{V. G. Ibarra-Sierra}
\author{J. C. Sandoval-Santana}
\address{Departamento de F\'isica, Universidad Aut\'onoma Metropolitana
Iztapalapa, Av. San Rafael Atlixco 186, Col. Vicentina,
09340 M\'exico D.F., Mexico} 
\author{J.L. Cardoso}
\author{A. Kunold}
\address{\'Area de F\'isica Te\'orica y Materia Condensada,
Universidad Aut\'onoma Metropolitana  Azcapotzalco,
Av. San Pablo 180, Col. Reynosa-Tamaulipas, Azcapotzalco,
 02200 M\'exico D.F., M\'exico }

 \date{\today}

 \begin{abstract}
 
We develop a Lie algebraic approach to systematically calculate
the evolution operator of the generalized two-dimensional
quadratic Hamiltonian with time-dependent coefficients.
Although the development of the Lie algebraic approach presented here is
mainly motivated by the two-dimensional quadratic Hamiltonian,
it may be applied to investigate the evolution operators of
any Hamiltonian having a dynamical algebra with a large number of elements.
We illustrate the method by finding the propagator and the Heisenberg
picture position and momentum operators for a two-dimensional
charge subject to uniform and constant electro-magnetic fields.
 \end{abstract}


 \maketitle 

\section{Introduction} \label{Introduction}
In many applications as radio-frequency ion traps 
\cite{PhysRevA.84.062104, PhysRevA.89.035401, PhysRevA.89.052332, 
PhysRevLett.66.527, PhysRevA.49.421, RevModPhys.75.281, PhysRevA.89.032502, 
PhysRevA.81.033402},  quantum optics \cite{PhysRevA.46.5885,1464-4266-4-3-379, 
Singh20104685, Mandal2004308}, cosmology \cite{PhysRevD.49.788,Pedrosa2007384},
quantum field theory \cite{Vergel20091360}, quantum 
dissipation \cite{caldirola:393, kanai:440, 
PhysRev.38.815,Um200263, 0305-4470-19-15-024, PhysRevA.68.052108, 
IbarraSierra201386}, magneto transport in lateral heterostructures
\cite{PhysRevB.88.245409,lei:233711,Inarrea201410,Kunold201378}
and even gravitational waves \cite{1751-8121-42-5-055307}
the time evolution of particles
in quadratic potentials is frequently examined.
The one-dimensional, generalized time-dependent quadratic Hamiltonian is
given by
\begin{equation}
\hat H=a_1(t) +a_2(t) \hat x+a_3(t) \hat p+a_4(t)\hat x^2+a_5(t)\hat p^2
+a_6(t)\left(\hat x\hat p+\hat p\hat x\right),\label{oneDQH}
\end{equation}
where $\hat x$ and $\hat p$ are the usual position and momentum operators
following the standard commutation relation $\left[\hat x,\hat p\right]=i\hbar$.
Aside from the simple harmonic oscillator, a large number of interesting systems 
arise from this Hamiltonian
as the linear potential \cite{PhysRevA.63.034102,PhysRevA.68.016101},
the driven harmonic oscillator \cite{1751-8121-43-38-385204,QUA:QUA23253},
Kanai-Caldirola Hamiltonians \cite{caldirola:393, kanai:440, 
PhysRev.38.815,Um200263, 0305-4470-19-15-024, PhysRevA.68.052108},
and time dependent harmonic oscillators i.e. an oscillator with
time-varying frequency \cite{:/content/aip/journal/jmp/55/11/10.1063/1.4901753,PhysRevA.68.052108,
:/content/aip/journal/jmp/56/3/10.1063/1.4914337}.
The time evolution generated by the
most general version of (\ref{oneDQH}) has been studied by means
of Lewis and Riesenfeld \cite{lewis:1458} invariants \cite{LopesdeLima20082253}
and through linear invariants \cite{PhysRevA.84.062104}.

Combining two one-dimensional generalized quadratic Hamiltonians
along the $x$ and $y$ coordinates and adding
cross terms for the position and momentum operators one arrives
to the most general form of the two-dimensional quadratic Hamiltonian
\begin{multline}
\hat H=a_1(t) +a_2(t) \hat x+a_3(t)\hat y
  +a_4(t)\hat{p}_x+a_5(t)\hat p_y
  +a_6(t)\hat x^2+a_7(t)\hat y^2
  +a_{8}(t)\hat x\hat y
  +a_9(t)\hat p_x^2\\
  +a_{10}(t)\hat p_y^2
  +a_{11}(t)\hat p_x\hat p_y
  +a_{12}(t)\left(\hat x\hat p_x+\hat p_x\hat x\right)
  +a_{13}(t)\left(\hat y\hat p_y+\hat p_y\hat y\right)\\
  +a_{14}(t)\hat x\hat p_y
  +a_{15}(t)\hat y\hat p_x,\label{twoDQH}
\end{multline}
where newly the position and momentum operators along the $x$ and $y$ axes
follow the standard commutation
relations $\left[\hat x,\hat y\right]=\left[\hat p_x,\hat p_y\right]=0$ and
$\left[\hat x,\hat p_x\right]=\left[\hat y,\hat p_y\right]=i\hbar$.
Hamiltonians as the one of a charged particle subject
to variable electromagnetic fields, two coupled one-dimensional oscillators
or the two dimensional harmonic oscillator stem from
this Hamiltonian. In particular, the Hamiltonian of a one-dimensional
generalized harmonic oscillator arises from (\ref{twoDQH}).

Some special cases emerging from these Hamiltonians
have been studied by diverse mathematical methods.
For instance the time dependent linear potential has been treated through 
the Lewis and Riesenfeld \cite{lewis:1458} invariant theory
\cite{PhysRevA.63.034102,PhysRevA.68.016101,maamache:1063},
Feynman's path integrals \cite{kiyoto:777,Khandekar1978, Feynman1948,
Feynman1950, Merzbarcher3th}, time-space 
transformation methods \cite{chao:981} and others \cite{PhysRevA.64.034101,
PhysRevA.71.014101}.
The quantum oscillator with time-dependent mass and frequency
has been dealt through the group-theoretical approach\cite{PhysRevA.44.2057},
unitary transformations \cite{PhysRevLett.66.527},
the Lewis and Riesenfeld 
invariant theory \cite{PhysRevA.55.3219,PhysRevA.20.550,0305-4470-34-37-321}.

Even though the most general form
of the one-dimensional quadratic Hamiltonian (\ref{oneDQH})
has been treated through the Lewis Riesenfeld theory
\cite{LopesdeLima20082253,PhysRevA.55.4023,PhysRevA.68.052108}
and linear invariants \cite{PhysRevA.84.062104},
the two-dimensional quadratic Hamiltonian (twoDQH) has only been studied for
a limited number of special cases. One of these corresponds to
a charged particle subject to a constant
uniform magnetic field and a quadratic potential \cite{0305-4470-17-4-022}
whose propagator was calculated by means of the path integral method.
The isotropic harmonic oscillator in the presence of a time dependent magnetic
field was investigated through the
unitary transformation approach\cite{1402-4896-73-6-024,IbarraSierra201386}.
The Lewis and Riesenfeld invariant
theory\cite{lewis:1458,PhysRevA.66.024103,PhysRevA.73.016101}
and quadratic invariants  \cite{contentaipjournaljmp52810.10631.3615516}
were applied to the study of
a charged particle subject to time-varying
electromagnetic fields \cite{Inarrea201410}.

Therefore, even though a wide variety of systems stemming from the Hamiltonian
in Eq. (\ref{twoDQH}) have been studied by diverse methods,
the evolution of the two-dimensional
generalized quadratic Hamiltonian's most general case 
has not been treated by any method to the extent of our knowledge.

The aim of this paper is therefore to develop a systematic method based
on the Lie algebraic approach mainly with the purpose of obtaining
the evolution operator of the two-dimensional, generalized
time-dependent quadratic Hamiltonian presented in (\ref{oneDQH}) .
Although most of this paper is devoted to the Lie algebra of
(\ref{twoDQH}), that consists of 15 generators, the presentation on
the Lie algebraic approach is general enough
to be applied to systems whose Hamiltonians  can be expanded in terms of
an arbitrarily  large number of generators.

This paper is organized as follows. In Section \ref{liealap}
we develop the Lie algebraic approach for a rather general
Hamiltonian consisting of the linear combination of an arbitrary
number of generators.
The Lie algebraic approach is applied to the generalized
two-dimensional Harmonic oscillator
in Section \ref{gqh}.
The general method is illustrated through the
example of a two-dimensional charged particle subject to an in-plane electric
field and a perpendicular magnetic field  in Section \ref{twodimcharge} .
In Section \ref{conclusions} we summarise and give general conclusions.

\section{The Lie algebraic approach}\label{liealap}
The Lie algebraic approach relies on the existence
of a set of $n$ operators
$\left\{\hat h_1, \hat h_2, \dots \hat h_n\right\}$
that form a closed Lie algebra $\mathcal{L}_n$.
This means that the commutator of any two elements of
$\mathcal{L}_n$ should be expressible as a linear combination of
its own elements
\begin{equation}
\left[\hat h_i,\hat h_j\right]=i\hbar\sum_{k=1}^nc_{i,j,k}\hat h_k,
\label{lie:commu}
\end{equation}
where $c_{i,j,k}$ are named structure constants of the algebra and contain
all of the information concerning the unitary group.

The Hamiltonian $\hat H$ of a given system is said to have a dynamical
algebra if it can be expressed as the linear combination
of the elements of $\mathcal L_n$
\begin{equation}
\hat H=\sum_{k=1}^n a_k\left(t\right)\hat h_k=\mathbf{\hat h}^{\top} \mathbf{a} ,
\end{equation}
where for the sake of simplicity we have defined the vectors
\begin{eqnarray}
\mathbf{\hat h}^{\top}&=& \left(\hat h_1,\hat h_2,\dots,\hat h_n \right),\\
\mathbf{a}^{\top} &=& \left(a_1,a_2,\dots,a_n\right).
\end{eqnarray}
The coefficients $a_i$ may in general be functions of time.
The key element behind the Lie algebraic approach is that
the general form of the evolution operator
of such type of Hamiltonian can be expressed as
\cite{CPA:CPA3160070404,jmathphys.1.1703993,
PhysRevA.18.89,0305-4470-21-22-015,PhysRevA.87.022116}
\begin{equation}
\hat{\mathcal{U}}=\hat U^\dagger=\exp\left[-i\hbar\sum_{i=1}^n\gamma_k(t)\hat h_k\right]
=\prod_{k=1}^n\hat U_k^\dagger
=\prod_{k=1}^n\exp\left[-i\hbar\alpha_k\left(t\right) \hat h_k\right]
\label{evolop}
\end{equation}
where $\hat U$ is an auxiliary unitary operator and
\begin{equation}
\hat U_k=\exp\left[i\alpha_k\left(t\right) \hat h_k\right], \,\,\,\,\,\,\,\,\,
k=1,\dots n,\label{unit:trans}
\end{equation}
are the elements of unitary group generated by $\hat h_k$
with transformation parameters $\alpha_k$.
As a direct consequence of
the algebra closure, any transformation
$\hat U_i$ acting on any generator  $\hat h_j$ yields
the linear combination of the same generators
\begin{equation}
\hat U_i\hat h_j\hat U_i^\dagger
=\sum_{k=1}^n\left(\mathcal{M}_i\right)_{j,k}\hat h_k
\,\,\,\,\, i,j,k=1, \dots n .
\label{transrules:gen}
\end{equation}
This expression can also be conveniently expressed as
\begin{equation}
\hat U_i\left(\alpha_i\right)\mathbf{\hat h}\hat U_i^\dagger\left(\alpha_i\right)\
=\mathcal{M}_i\left(\alpha_i\right)\mathbf{\hat h}.\label{vectran}
\end{equation}
These represent the transformation rules of $\mathcal{L}_n$ that
are completely determined by the $\mathcal{M}_i$ matrices.

Additionally, referring to (\ref{re:cuan}), we see that all of the
transformations given above
acting on the energy operator yield
\begin{equation}
\hat U_i \hat p_t\hat U_i^\dagger=\hat p_t+\dot\alpha_i(t)\hat h_i
=\hat p_t+\mathbf{\hat h}^\top \mathcal{I}_i\boldsymbol{\dot \alpha},
\,\,\,\,\, i=1, \dots n.\label{enerrules:gen}
\end{equation}
where $\left(\mathcal{I}_i\right)_{jk}=\delta_{i,j}\delta_{i,k}$,
\begin{equation}
\boldsymbol{\alpha^\top}=\left(\alpha_1,\alpha_2,\dots,\alpha_n\right)
\end{equation}
and the overdot denotes the time derivative thus
\begin{equation}
\boldsymbol{\dot{\alpha}^\top}=\left(\frac{d\alpha_1}{dt},\frac{d\alpha_2}{dt},
\dots,\frac{d\alpha_n}{dt}\right).
\end{equation}

Let us now proceed to finding the evolution operator.
The transformation parameters $\alpha_i$
are in general time-dependent functions yet to be found.
Once these functions are known, the evolution operator
is completely determined
as can be seen from Eq. (\ref{evolop}). However, calculating them
is not an easy task. 
To do so, let us first consider 
Schr\"odinger equation
\begin{equation}\label{ec.schrodinger}
\hat{H} \left\vert \psi \left( t\right) \right\rangle =\hat{p}_t \left\vert\
\psi\left(t\right)\right\rangle,
\end{equation}
where $\hat p_t=i\hbar\partial/\partial t$ is the energy operator.
It is convenient to introduce the Floquet operator\cite{Heinzpeter2006}
\begin{equation}\label{floquet}
\hat{\mathcal H} =\hat{H} -\hat{p}_t,
\end{equation}
since it allows to express Schr\"odinger equation in the compact form  
\begin{equation}\label{shro}
\hat{\mathcal H} \left\vert \psi \left(t \right) \right\rangle= 0.
\end{equation}

Let us now assume that a set of unitary transformation parameters
$\alpha_1, \alpha_2, \dots \alpha_n$, exists
such that if $\hat U=\hat U_n\dots \hat U_2\hat U_1$ is applied to
the Shr\"odinger eqution (\ref{shro}), the Floquet operator is reduced
to the energy operator $\hat p_t$, namely
\begin{equation}
\hat{U} \hat{\mathcal H} \hat{U}^{\dagger} \hat{U} \left\vert \psi \left( 
t\right)\right\rangle = -\hat{p}_t \left[\hat{U} \left\vert \psi \left( 
t\right)\right\rangle\right]=0.\label{prescription}
\end{equation}

Reminding that $\hat p_t$ is $\hbar$ times a time derivative, it is clear
that $\hat{U}\left\vert\psi\left(t\right)\right\rangle$ must be
a time-independent ket, say
\begin{equation}
\hat{U} \left\vert \psi \left( t\right) \right\rangle = \left\vert \psi \left(
0\right) \right\rangle,
\end{equation}
or equivalently
\begin{equation}
\left\vert \psi \left( t\right) \right\rangle = \hat{U}^{\dagger} \left\vert 
\psi \left( 0\right) \right\rangle.\label{iamevolve}
\end{equation}
According to the considerations above this equation states that
\begin{equation}
\hat{\mathcal{U}}=\hat{U}^{\dagger}
=\hat U_1^\dagger\hat U_2^\dagger \dots \hat U_n^\dagger,
\label{theevolutionoperator}
\end{equation}
is in fact the time evolution operator (\ref{evolop}).
Hence Eq. (\ref{prescription}) gives us a prescription for finding
the evolution operator: if the transformation $\hat U$ reduces
the Floquet operator $\hat{\mathcal H}$ to the energy operator
$\hat p_t$ then $\hat{\mathcal U}=\hat U^\dagger$ is the evolution operator.

At this point it is clear that in order to calculate any operator
in the Heisenberg picture we can successively apply $\hat U_1$ to
$\hat U_n$. Thus, the Heisenberg picture operator of $\mathbf{\hat h}$ is given by
\begin{equation}
\mathbf{\hat h}_{H}=\hat U_n\dots\hat U_2\hat U_1 \mathbf{\hat h}
\hat U_1^\dag\hat U_2^\dag\dots\hat U_n^\dag
=\mathcal{M}_1 \mathcal{M}_2\dots\mathcal{M}_n \mathbf{\hat h}.
\label{gen:heispict}
\end{equation}
This expression is easily evaluated by using the transformation rules 
(\ref{vectran}).
Moreover, if $A(\mathbf{\hat h})$ is an analytic function of
the generators $\mathbf{\hat h^{\top}}=(\hat h_1,\hat h_2,\dots\hat h_n)$ the Heisenberg
picture of $A$ is given by
\begin{eqnarray}
A_H(\mathbf{\hat h}) = \hat U_n\dots\hat U_2\hat U_1 A(\mathbf{\hat h})&&
\hat U_1^\dag\hat U_2^\dag\dots \hat U_n^\dag\nonumber\\
 &&= A(\mathbf{\hat h}_H)=A(\hat h_{H1},\hat h_{H2},\dots\hat h_{Hn}).
 \label{gen:heispictfun}
\end{eqnarray}

Even though (\ref{theevolutionoperator}) gives the general form of the
evolution operator,
we have not yet established the relation between the transformation
parameters $\boldsymbol{\alpha}$ and the Hamiltonian coefficients
 $\boldsymbol{a}$ that insure
that condition (\ref{prescription}) is met. In order to acomplish
this, we use Eqs. (\ref{transrules:gen}) and (\ref{enerrules:gen})
and infer that the general structure of the transformed
Floquet operator must be
\begin{equation}
\hat{U} \hat{\mathcal H} \hat{U}^{\dagger}=
\sum_{i=1}^n
u_i\left(\mathbf{a}, \boldsymbol{\alpha},\boldsymbol{\dot\alpha}\right)
\hat h_i-\hat p_t=\mathbf{\hat h}^\top \mathbf{u} -\hat p_t.\label{trans:floquet}
\end{equation}
Furthermore, according to Eqs. (\ref{vectran})
and (\ref{enerrules:gen}), $\mathbf{u}$ must be a linear function of
$\boldsymbol{\dot\alpha}$ of the form
\begin{equation}
\mathbf{u}\left(\mathbf{a}, \boldsymbol{\alpha},\boldsymbol{\dot\alpha}\right)
= \boldsymbol{w} \left(\mathbf{a},\boldsymbol{\alpha}\right)+ \nu\left( 
\boldsymbol{\alpha}\right) \boldsymbol{\dot{\alpha}},
\label{uiform}
\end{equation}
where
\begin{eqnarray}
\boldsymbol{w} \left(\mathbf{a},\boldsymbol{\alpha}\right) &=& 
\mathcal{M}_n^\top\left(\alpha_n\right)
\dots \mathcal{M}_2^\top\left(\alpha_2\right)
\mathcal{M}_1^\top\left(\alpha_1\right) \mathbf{a},
\nonumber \\
\nu\left(\boldsymbol{\alpha}\right) &=& \mathcal{M}_n^\top\left(\alpha_n\right)
\dots \mathcal{M}_3^\top\left(\alpha_3\right)
\mathcal{M}_2^\top\left(\alpha_2\right)\mathcal{I}_1
+\mathcal{M}_n^\top\left(\alpha_n\right)
\dots \mathcal{M}_4^\top\left(\alpha_4\right)
\mathcal{M}_3^\top\left(\alpha_3\right)\mathcal{I}_2
\nonumber \\
&&\dots+\mathcal{M}_n^\top\left(\alpha_n\right)
\mathcal{I}_{n-1}+\mathcal{I}_{n}. 
\end{eqnarray}
In order for (\ref{trans:floquet}) to reduce to the energy operator $\hat p_t$
the $u_i$ coefficients must vanish giving rise to the following 
system of $n$ ordinary coupled differential equations
\begin{equation}
\mathbf{u}\left(\mathbf{a}, \boldsymbol{\alpha},\boldsymbol{\dot\alpha}\right)=0.\label{ode:sys}
\end{equation}
These equations provide the means to establish the explicit form of
the transformation parameters that fulfil condition (\ref{prescription}).
Although in principle (\ref{ode:sys}) would suffice to
determine the transformation parameters $\boldsymbol{\alpha}$, 
algebras formed by a large number of operators yield very complex
system of ordinary differential equations hindering their solution.
Notwithstanding
it is possible to simplify the $u_i$ coefficients even further
into the linear combination
\begin{equation}
\mathbf{u}\left(\mathbf{a}, \boldsymbol{\alpha},\boldsymbol{\dot\alpha}\right)
=\nu\left(\boldsymbol{\alpha}\right)\boldsymbol{\mathcal{E}}\left(\mathbf{a}, 
\boldsymbol{\alpha},\boldsymbol{\dot\alpha}\right),
\label{uiformtwo}
\end{equation}
where the elements of the vector
$\boldsymbol{\mathcal E}^{\top}=\left(\mathcal{E}_1,\mathcal{E}_2,\dots,\mathcal{E}_n\right)$
are more simple differential equations of the form
\begin{equation}
\boldsymbol{\mathcal{E}}\left(\mathbf{a}, 
\boldsymbol{\alpha},\boldsymbol{\dot\alpha}\right)=\boldsymbol{\mu}\left(\mathbf{a},
\boldsymbol{\alpha}\right)- \boldsymbol{\dot { \alpha }} =0.
\label{simpleodes}
\end{equation}
Even though from the above expressions it is evident that
$\boldsymbol{\mu}\left(\mathbf{a},\boldsymbol{\alpha}\right)=\nu^{-1}
\left(\boldsymbol{\alpha}\right) \boldsymbol{w} \left(\mathbf{a} , \boldsymbol{\alpha}\right)$,
$\boldsymbol{w} \left(\mathbf{a} , \boldsymbol{\alpha}\right)$ is not essential to know
the explicit form of the coefficients $u_i$
as functions of $\mathbf{a}$, $\boldsymbol{\alpha}$ and $\boldsymbol{\dot\alpha}$; it suffices to work
them out from (\ref{trans:floquet})
by successively applying the transformation rules (\ref{vectran})
to the Floquet operator. Thereby, from Eqs. (\ref{uiform}) and (\ref{uiformtwo})
\begin{equation}
\nu_{i,j}\left(\mathbf{a},\boldsymbol{\alpha}\right)=\frac{\partial u_i}{\partial \dot{\alpha}_j}
=-\frac{\partial u_i}{\partial \mathcal{E}_j},\label{nudef}
\end{equation} 
and since $u$ might be expressed either
as a linear combination of $\dot{\boldsymbol{\alpha}}$ or $\boldsymbol{\mathcal{E}}$
the equations of the form (\ref{simpleodes}) may be obtained from
\begin{equation}
\boldsymbol{\mathcal{E}}\left(\mathbf{a}, 
\boldsymbol{\alpha},\boldsymbol{\dot\alpha}\right)
=\nu^{-1}\left(\boldsymbol{\alpha}\right)\mathbf{u}\left(\mathbf{a}, 
\boldsymbol{\alpha},\boldsymbol{\dot\alpha}\right)=0,
\label{equations}
\end{equation}
provided that $\det \nu\ne 0$.
In order to know
the evolution operator i.e. the transformation parameters' explicit form one must
find the solution to the system of ordinary differential equations (\ref{equations}).

To summarise we can reduce the method into five steps: a) The Floquet operator is transformed
by applying the whole set of unitary transformations generated by $\mathcal{L}_n$. b)
Identify the $u_i$ coefficients from the transformed Floquet operator. c) Derive the $\nu$ matrix
through Eq. (\ref{nudef}). d) Obtain the simplified set of ordinary equations by using Eq. (\ref{equations}).
e) Solve the set of ordinary differential equations for the $\boldsymbol{\alpha}$ parameters. f) The evolution
operator is finally obtained by plugging this solution into the general form of the evolution
operator in Eq. (\ref{evolop}).

Finally, the Green function may be obtained
by splitting the evolution operator's matrix element into the
the ones concerning each of the $n$ unitary transformations
as
\begin{multline}
G(x,y,t; x^\prime, y^\prime,0)=
 \left\langle x,y\left\vert\hat{U}^\dag\left(t\right)\right\vert x^\prime,y^\prime \right\rangle\\
 =\int dx_1 dy_1\int dx_2 dy_2\dots\int dx_{n-1}dy_{n-1}
 \left\langle x,y \left\vert\hat{U}_1^\dag\left(t\right)\right\vert x_1, y_1 \right\rangle
 \left\langle x_1,y_1\left\vert\hat{U}_2^\dag\left(t\right)\right\vert x_2,y_2 \right\rangle\\
\times \dots\left\langle x_{n-2},y_{n-2}\left\vert\hat{U}_{n-1}^\dag\left(t\right)\right\vert 
 x_{n-1},y_{n-1}\right\rangle
 \left\langle x_{n-1},y_{n-1}\left\vert\hat{U}_{n}^\dag\left(t\right)\right\vert 
 x^\prime,y^\prime\right\rangle.\label{greenfunc}
\end{multline}
The matrix elements of $\hat U_1^\dag$, $\hat U_2^\dag$, $\dots$
$\hat U_n^\dag$ are readily calculated by
using the transformation rules.

\section{Generalized two-dimensional quadratic Hamiltonians}\label{gqh}

In this section we develop the Lie algebraic approach to obtain the
evolution operator of the general two-dimensional quadratic Hamiltonian.
To motivate the discussion let us take the Hamiltonian
of a two-dimensional charged particle in perpendicular magnetic field and
in-plane electric fields as the starting point.
This Hamiltonian is given by
\begin{equation}
\hat H=\frac{1}{2m}\left(\hat p_x^2+\hat p_y^2\right)
+\frac{e^2B^2}{8m}\left(\hat x^2+\hat y^2\right)
+\frac{eB}{2m}\left(\hat x\hat p_y-\hat y\hat p_x\right)
+eE_x\hat x+eE_y\hat y,
\label{gfho}
\end{equation}
where $m$ and $q=-e$ are the particle's mass and charge, $B$, $E_x$ and $E_y$
are the perpendicular magnetic and in-plane electric field components.
The scalar and vector potentials
are expressed in the symmetric gauge as $\phi=-E_x(t)\hat x-E_y(t)\hat y$,
$A_x=-B\hat y/2$ and $A_y=B\hat x/2$.
The position and momentum operators $\hat x$, $\hat y$, $\hat p_x$
and $\hat p_y$ fulfil the usual commutation relations
$\left[\hat x,\hat p_x\right]=\left[\hat y,\hat p_y\right]=i\hbar$ and
$\left[\hat x,\hat y\right]=\left[\hat p_x,\hat p_y\right]
=\left[\hat x,\hat p_y\right]=\left[\hat y,\hat p_x\right]=0$.

In principle this Hamiltonian is expressed as a linear combination
of $\hat x$, $\hat y$, $\hat x^2$, $\hat y^2$,
$\hat p_x^2$, $\hat p_y^2$, $\hat x\hat p_y$ and $\hat y\hat p_x$
with time dependent coefficients.
However, these eight operators
alone do not form a closed Lie algebra under commutation given that,
for example, the commutators $\left[\hat x,\hat p_x\right]=i\hbar \hat 1$,
$\left[\hat x,\hat p_x^2\right]=2i\hbar \hat p_x$ and
$\left[\hat x^2,\hat p_x^2\right]=2i\hbar\left(\hat x \hat p+\hat p\hat x\right)$
yield operators outside the original set, namely $\hat 1$,
$\hat p_x$ and $\hat x \hat p_x+\hat p_x\hat x$. Then there follows that, in order to close
the algebra, the set must be extended to
\begin{align}
 \hat h_1 &=\hat 1, & \hat h_2 &=\hat x, & \hat h_3 &=\hat y,  & \hat h_4 &=\hat p_x,
    & \hat h_5 &=\hat p_y , \nonumber \\
 \hat h_6 &=\hat x^2, & \hat h_7 &=\hat y^2, & \hat h_8 &=\hat x\hat y, & \hat h_9 &=\hat p_x^2, &
 \hat h_{10} &=\hat p_y^2,   \label{fho:base:1} \\
 \hat h_{11} &=\hat p_x\hat p_y, & \hat h_{12} &=\hat x\hat p_x+\hat p_x\hat x, & 
 \hat h_{13} &=\hat y\hat p_y+\hat p_y\hat y,&  \hat h_{14} &=\hat x \hat p_y, &
 \hat h_{15}&=\hat y\hat p_x.\nonumber
\end{align}
Indeed, the commutation relations for these operators yield
a closed algebra summarised in Table \ref{ta:commutators}.
We henceforth call this
algebra $\mathcal{L}_{15}$. These commutation relations can be comprehended as a particular
realization of (\ref{lie:commu}) where the structure constants are related to the
coefficients found in Table  \ref{ta:commutators} .
Also in this Table, the particular generator ordering in Eq. (\ref{fho:base:1}) reveals
the two following different sub-algebras (enclosed in squares)
$\left\{\hat h_1, \hat h_2, \hat h_3, \hat h_4, \hat h_5\right\}$ and
$\left\{\hat h_1, \hat h_2, \hat h_3, \hat h_4, \hat h_5\, \hat h_6, \hat h_7, \hat h_8 \right\}$.
Not as evident as the previous ones, 
one can find even more sub-algebras in $\mathcal{L}_{15}$ that correspond to
relevant physical problems.
For example
$\left\{\hat h_1, \hat h_2, \hat h_4 \right\}$ and $\left\{\hat h_1, \hat h_3, \hat h_5 \right\}$ 
are also sub-algebras of $\mathcal{L}_{15}$. In particular
$\left\{\hat h_6, \hat h_9, \hat h_{12} \right\}$ or $\left\{\hat h_7, \hat h_{10}, \hat h_{13} \right\}$
form the $SU(1,1)$ Lie algebra that has been used to study Kanai-Caldirola
Hamiltonians through the Lie algebraic approach \cite{0305-4470-21-22-015,1751-8121-42-5-055307}.
The sub-algebras $\left\{\hat h_1, \hat h_2, \hat h_4,  \hat h_6, \hat h_9, \hat h_{12}\right\}$ or
$\left\{\hat h_1, \hat h_3, \hat h_5,  \hat h_7, \hat h_{10}, \hat h_{13}\right\}$
correspond to the generalised one-dimensional
harmonic oscillator\cite{LopesdeLima20082253,PhysRevA.84.062104} along the
$x$ and $y$ axis respectively.

One can easily express  Hamiltonian (\ref{gfho})
as a linear combination of the $\mathcal{L}_{15}$ elements 
\begin{multline}
\hat H=a_1\hat h_1+a_2\hat h_2+a_3\hat h_3+a_4\hat h_4+a_5\hat h_5+a_6\hat h_6
            +a_7\hat h_7 +a_8\hat h_8+a_9\hat h_9+a_{10}\hat h_{10}+a_{11}\hat h_{11}\\
            +a_{12}\hat h_{12}+a_{13}\hat h_{13}+a_{14}\hat h_{14}+a_{15}\hat h_{15},
\label{ham:gao}
\end{multline}
where $a_2=e E_x$, $a_3=eE_y$
$a_6=a_7=e^2B^2/8m$, $a_9=a_{10}=1/2m$, $a_{14}=-a_{15}=eB/2m$
and $a_1=a_4=a_5=a_8=a_{11}=a_{12}=a_{13}=0$.
In the most general case, when the coefficients $a_1$ to $a_{15}$
are non-vanishing functions of time, we call (\ref{ham:gao}) the
generalized two-dimensional quadratic Hamiltonian.
Many Hamiltonians of physical significance arise from (\ref{ham:gao})
for example, a single electron in an elliptically shaped quantum dot
with quadratic confining potential, an electron subject to variable
electromagnetic field or two-dimensional quadratic Kanai-Caldirola 
Hamiltonians among others.
  
\begin{table}
\caption{\label{ta:commutators} Commutation rules of the
generalized two-dimensionall harmonic oscillator. The complete set of operators is given 
by $\hat h_1=\hat 1$, $\hat h_2=\hat x$, $\hat h_3=\hat y$ , $\hat h_4= \hat p_x$,  
$\hat h_5= \hat p_y$, $\hat h_6=\hat x^2$, $\hat h_7=\hat y^2$, 
$\hat h_8=\hat x \hat y$, $\hat h_9=\hat p_x^2$, $\hat h_{10}=\hat p_y^2$, 
$\hat h_{11}=\hat p_x \hat p_y$, $\hat h_{12}=\hat x \hat p_x + \hat p_x \hat x$, 
$\hat h_{13}=\hat y \hat p_y + \hat p_y \hat y$, $\hat h_{14}=\hat x \hat p_y$
and $\hat h_{15}= \hat y \hat p_x$. The sub-algebras 
$\left\{\hat h_1, \hat h_2, \hat h_3, \hat h_4, \hat h_5\right\}$ and
$\left\{\hat h_1, \hat h_2, \hat h_3, \hat h_4, \hat h_5\, \hat h_6, \hat h_7, \hat h_8 \right\}$
are enclosed in the squares.}
\begin{center}
\resizebox*{1.0\textwidth}{!}{
\begin{tabular}{c|ccccccccccccccc|}
 & $\hat h_1]$ & $\hat h_2]$ & $\hat h_3]$ & $\hat h_4]$ & 
\multicolumn{1}{c|}{$\hat h_5]$} & $\hat h_6]$ & $\hat h_7]$ & 
\multicolumn{1}{c|}{$\hat h_8]$} & $\hat h_9]$ & $\hat h_{10}]$ &
$\hat h_{11}]$ & $\hat h_{12}]$ & $\hat h_{13}]$ & $\hat h_{14}]$ & 
$\hat h_{15}]$ \\
\hline
$\frac{1}{i\hbar} [\hat h_1,$ & $0$ & $0$ & $0$ & $0$ & 
\multicolumn{1}{c|}{$0$} & $0$ & $0$ & \multicolumn{1}{c|}{$0$} &
 $0$ & $0$ & $0$ & $0$ & $0$ & $0$ & $0$ \\
$\frac{1}{i\hbar} [\hat h_2,$ & $0$ & $0$ & $0$ & $\hat h_1$ & 
\multicolumn{1}{c|}{$0$} & $0$ & $0$ & \multicolumn{1}{c|}{$0$} & 
$2 \hat h_4$ & $0$ & $\hat h_5$ & $2\hat h_2$ & $0$ & $0$ & 
$\hat h_3$ \\
$\frac{1}{i\hbar} [\hat h_3,$ & $0$ & $0$ & $0$ & $0$ & 
\multicolumn{1}{c|}{$\hat h_1$} & $0$ & $0$ & \multicolumn{1}{c|}{$0$} & $0$ 
& $2\hat h_5$ & $\hat h_4$ & $0$ & $2\hat h_3$ & $\hat h_2$ & $0$ \\
$\frac{1}{i\hbar} [\hat h_4,$ & $0$ & $-\hat h_1$ & $0$ & $0$ & 
\multicolumn{1}{c|}{$0$} & $-2\hat h_2$ & $0$ & 
\multicolumn{1}{c|}{$-\hat h_3$} & $0$ & $0$ & $0$ & $-2\hat h_4$ & $0$ & 
$-\hat h_5$ & $0$ \\
$\frac{1}{i\hbar} [\hat h_5,$ & $0$ & $0$ & $-\hat h_1$ & $0$ &
\multicolumn{1}{c|}{$0$} & $0$ & $-2\hat h_3$ & 
\multicolumn{1}{c|}{$-\hat h_2$} & $0$ & $0$ & $0$ & $0$ & $-2\hat h_5$ & 
$0$ & $-\hat h_4$ \\
\cline{1-6}
$\frac{1}{i\hbar} [\hat h_6,$ & $0$ & $0$ & $0$ & $2\hat h_2$ & $0$ & $0$ & 
$0$ & \multicolumn{1}{c|}{$0$} & $2\hat h_{12}$ & $0$ & $2\hat h_{14}$ & 
$4\hat h_6$ & $0$ & $0$ & $2\hat h_8$ \\
$\frac{1}{i\hbar} [\hat h_7,$ & $0$ & $0$ & $0$ & $0$ & $2 \hat h_3$ & $0$ & 
$0$ & \multicolumn{1}{c|}{$0$} & $0$ & $2\hat h_{13}$ & $2\hat h_{15}$ & $0$ & 
$4\hat h_7$ & $2\hat h_8$ & $0$ \\
$\frac{1}{i\hbar} [\hat h_8,$ & $0$ & $0$ & $0$ & $\hat h_3$ & $\hat h_2$ & 
$0$ & $0$ & \multicolumn{1}{c|}{$0$} & $\hat h_{15}$ & $2\hat h_{14}$ & 
$\frac{\hat h_{12} +\hat h_{13}}{2}$ & $2\hat h_8$ & 
$2\hat h_8$ & $\hat h_6$ & $\hat h_7$ \\
\cline{1-9}
$\frac{1}{i\hbar} [\hat h_9,$ & $0$ & $-2\hat h_4$ & $0$ & $0$ & $0$ & 
$-2\hat h_{12}$ & $0$ & $-\hat h_{15}$ & $0$ & $0$ & $0$ & $-4\hat h_9$ & 
$0$ & $-2\hat h_{11}$ & $0$ \\
$\frac{1}{i\hbar} [\hat h_{10},$ & $0$ & $0$ & $-2\hat h_5$ & $0$ & $0$ & $0$ & 
$-2\hat h_{13}$ & $-2\hat h_{14}$ & $0$ & $0$ & $0$ & $0$ & $-4\hat h_{10}$ & 
$0$ & $-2\hat h_{11}$\\
$\frac{1}{i\hbar} [\hat h_{11},$ & $0$ & $-\hat h_5$ & $-\hat h_4$ & $0$ & 
$0$ & $-2\hat h_{14}$ & $-2\hat h_{15}$ & $-\frac{\hat h_{12}+\hat h_{13}}{2}$ & 
$0$ & $0$ & $0$ & $-2\hat h_{11}$ & $-2\hat h_{11}$ & $-\hat h_{10}$ & 
$-\hat h_9$ \\
$\frac{1}{i\hbar} [\hat h_{12},$ & $0$ & $-2\hat h_2$ & $0$ & $2\hat h_4$ & 
$0$ & $-4\hat h_6$ & $0$ & $-2\hat h_8$ & $4\hat h_9$ & $0$ & $2\hat h_{11}$ & 
$0$ & $0$ & $-2\hat h_{14}$ & $2\hat h_{15}$ \\
$\frac{1}{i\hbar} [\hat h_{13},$ & $0$ & $0$ & $-2\hat h_3$ & $0$ & $2\hat h_5$ &
$0$ & $-4\hat h_7$ & $-2\hat h_8$ & $0$ & $4\hat h_{10}$ & $2\hat h_{11}$ & 
$0$ & $0$ & $2\hat h_{14}$ & $-2\hat h_{15}$ \\
$\frac{1}{i\hbar} [\hat h_{14},$ & $0$ & $0$ & $-\hat h_2$ & $\hat h_5$ & $0$ & 
$0$ & $-2\hat h_8$ & $-\hat h_6$ & $2\hat h_{11}$ & $0$ & $\hat h_{10}$ & 
$2\hat h_{14}$ & $-2\hat h_{14}$ & $0$ & $-\frac{\hat h_{12} -\hat h_{13}}{2}$ \\
$\frac{1}{i\hbar} [\hat h_{15},$ & $0$ & $-\hat h_3$ & $0$ & $0$ & $\hat h_4$ & 
$-2\hat h_8$ & $0$ & $-\hat h_7$ & $0$ & $2\hat h_{11}$ & $\hat h_9$ & 
$-2\hat h_{15}$ & $2\hat h_{15}$ & $\frac{\hat h_{12} -\hat h_{13}}{2}$ & $0$ \\
\hline

\end{tabular}
}
\end{center}
\end{table}

\begin{table}
  \caption{\label{transf:1-8} Transformations rules for $\hat U_2$-$\hat U_8$}
  \begin{center}
    \resizebox*{1.0\textwidth}{!}{
    \begin{tabular}{c|cccccccc|}
      & $\hat U_2 \{ \bullet \} \hat U_2^\dagger$ &
      $\hat U_3 \{ \bullet \} \hat U_3^\dagger$ &
      $\hat U_4 \{ \bullet \} \hat U_4^\dagger$ &
      $\hat U_5 \{ \bullet \} \hat U_5^\dagger$ &
      $\hat U_6 \{ \bullet \} \hat U_6^\dagger$ &
      $\hat U_7 \{ \bullet \} \hat U_7^\dagger$ &
      $\hat U_8 \{ \bullet \} \hat U_8^\dagger$ \\
      \hline
      $\hat h_1$ & $\hat h_1$ & $\hat h_1$ & $\hat h_1$ &
      $\hat h_1$ & $\hat h_1$ & $\hat h_1$ & $\hat h_1$ \\
      $\hat h_2$ & $\hat h_2$ & $\hat h_2$ &
      $\hat h_2 +\alpha_4 \hat h_1$ & $\hat h_2$ & $\hat h_2$ & $\hat h_2$ &
      $\hat h_2$ \\
      $\hat h_3$ & $\hat h_3$ & $\hat h_3$ & $\hat h_3$ &
      $\hat h_3 +\alpha_5 \hat h_1$ & $\hat h_3$ & $\hat h_3$ & $\hat h_3$ \\
      $\hat h_4$ & $\hat h_4 -\alpha_2 \hat h_1$ & $\hat h_4$ &
      $\hat h_4$ & $\hat h_4$ & $\hat h_4 -2\alpha_6 \hat h_2$ & $\hat h_4$ &
      $\hat h_4 -\alpha_8 \hat h_3$ \\
      $\hat h_5$ & $\hat h_5$ & $\hat h_5 -\alpha_3 \hat h_1$ &
      $\hat h_5$ & $\hat h_5$ & $\hat h_5$ & $\hat h_5 -2\alpha_7 \hat h_3$ &
      $\hat h_5 -\alpha_8 \hat h_2$ \\
      $\hat h_6$ & $\hat h_6$ & $\hat h_6$ &
      $\hat h_6 +2\alpha_4 \hat h_2 +\alpha_4^2 \hat h_1$ & $\hat h_6$ &
      $\hat h_6$ & $\hat h_6$ & $\hat h_6$ \\
      $\hat h_7$ & $\hat h_7$ & $\hat h_7$ & $\hat h_7$ &
      $\hat h_7 +2\alpha_5 \hat h_3 +\alpha_5^2 \hat h_1$ & $\hat h_7$ &
      $\hat h_7$ & $\hat h_7$ \\
      $\hat h_8$ & $\hat h_8$ & $\hat h_8$ &
      $\hat h_8 +\alpha_4 \hat h_3$ & $\hat h_8 +\alpha_5 \hat h_2$ &
      $\hat h_8$ & $\hat h_8$ & $\hat h_8$ \\
      $\hat h_9$ &
      $\hat h_9 -2\alpha_2 \hat h_4 +\alpha_4^2 \hat h_1$ & $\hat h_9$ &
      $\hat h_9$ & $\hat h_9$ &
      $\hat h_9 -2\alpha_6 \hat h_{12} +4\alpha_6^2 \hat h_6$ & $\hat h_9$ &
      $\hat h_9 -2\alpha_8 \hat h_{15} +\alpha_8^2 \hat h_7$ \\
      $\hat h_{10}$ & $\hat h_{10}$ & 
      $\hat h_{10}-2\alpha_3 \hat h_5 +\alpha_3^2 \hat h_1$ &
      $\hat h_{10}$ & $\hat h_{10}$ & $\hat h_{10}$ &
      $\hat h_{10} -2\alpha_7 \hat h_{13} +4\alpha_7^2 \hat h_7$ &
      $\hat h_{10} -2\alpha_8 \hat h_{14} +\alpha_8^2 \hat h_6$ \\
      $\hat h_{11}$ & $\hat h_{11} -\alpha_2 \hat h_5$ &
      $\hat h_{11} -\alpha_3 \hat h_4$ & $\hat h_{11}$ & $\hat h_{11}$ &
      $\hat h_{11} -2\alpha_6 \hat h_{14}$ & $\hat h_{11} -2\alpha_7 \hat h_{15}$ &
      $\hat h_{11}-\alpha_8\frac{\hat h_{12}+\hat h_{13}}{2}+\alpha_8^2\hat h_8$\\
      $\hat h_{12}$ & $\hat h_{12} -2\alpha_2 \hat h_2$ &
      $\hat h_{12}$ & $\hat h_{12} +2\alpha_4 \hat h_4$ & $\hat h_{12}$ &
      $\hat h_{12} -4\alpha_6 \hat h_6$ & $\hat h_{12}$ &
      $\hat h_{12} -2\alpha_8 \hat h_8$ \\
      $\hat h_{13}$ & $\hat h_{13}$ &
      $\hat h_{13} -2\alpha_3 \hat h_{3}$ & $\hat h_{13}$ &
      $\hat h_{13} +2\alpha_5 \hat h_{5}$ & $\hat h_{13}$ &
      $\hat h_{13} -4\alpha_7 \hat h_{7}$ & $\hat h_{13} -2\alpha_8 \hat h_8$ \\
      $\hat h_{14}$ & $\hat h_{14}$ &
      $\hat h_{14} -\alpha_3 \hat h_2$ & $\hat h_{14} +\alpha_4 \hat h_5$ &
      $\hat h_{14}$ & $\hat h_{14}$ & $\hat h_{14} -2\alpha_7 \hat h_8$ &
      $\hat h_{14} -\alpha_8 \hat h_6$ \\
      $\hat h_{15}$ & $\hat h_{15} -\alpha_2 \hat h_3$ & 
      $\hat h_{15}$ & $\hat h_{15}$ & $\hat h_{15} +\alpha_5 \hat h_4$ &
      $\hat h_{15} -2\alpha_6 \hat h_8$ & $\hat h_{15}$ &
      $\hat h_{15} -\alpha_8 \hat h_7$ \\
      \hline
      $\hat p_t$ & $\hat p_t +\dot \alpha_{2} \hat h_{2}$ &
      $\hat p_t +\dot \alpha_{3} \hat h_{3}$ &
      $\hat p_t +\dot \alpha_{4} \hat h_{4}$ &
      $\hat p_t +\dot \alpha_{5} \hat h_{5}$ &
      $\hat p_t +\dot \alpha_{6} \hat h_{6}$ &
      $\hat p_t +\dot \alpha_{7} \hat h_{7}$ &
      $\hat p_t +\dot \alpha_{8} \hat h_{8}$
    \end{tabular}}
  \end{center}
\end{table}

\begin{table}
  \caption{\label{transf:9-15} Transformations rules for
    $\hat U_9$-$\hat U_{15}$}
  \begin{center}
    \resizebox*{1.0\textwidth}{!}{
    \begin{tabular}{c|cccccccc|}
       & $\hat U_9 \{ \bullet \} \hat U_9^\dagger$ &
      $\hat U_{10} \{ \bullet \} \hat U_{10}^\dagger$ &
      $\hat U_{11} \{ \bullet \} \hat U_{11}^\dagger$ &
      $\hat U_{12} \{ \bullet \} \hat U_{12}^\dagger$ &
      $\hat U_{13} \{ \bullet \} \hat U_{13}^\dagger$ &
      $\hat U_{14} \{ \bullet \} \hat U_{14}^\dagger$ &
      $\hat U_{15} \{ \bullet \} \hat U_{15}^\dagger$ \\
      \hline
      $\hat h_1$ & $\hat h_1$ & $\hat h_1$ & $\hat h_1$ &
      $\hat h_1$ & $\hat h_1$ & $\hat h_1$ & $\hat h_1$ \\
      $\hat h_2$ & $\hat h_2 +2\alpha_9 \hat h_4$ & $\hat h_2$ &
      $\hat h_2 +\alpha_{11} \hat h_5$ &
      $\hat h_2 \exp{\left(2\alpha_{12}\right)}$ &
      $\hat h_2$ & $\hat h_2$ & $\hat h_2 +\alpha_{15} \hat h_3$ \\
      $\hat h_3$ & $\hat h_3$ & $\hat h_3 +2\alpha_{10} \hat h_5$ &
      $\hat h_3 +\alpha_{11} \hat h_4$ & $\hat h_3$ &
      $\hat h_3 \exp{\left(2\alpha_{13}\right)}$ & 
      $\hat h_3 +\alpha_{14} \hat h_2$ & $\hat h_3$ \\
      $\hat h_4$ & $\hat h_4$ & $\hat h_4$ & $\hat h_4$ &
      $\hat h_4 \exp{\left(-2\alpha_{12}\right)}$ & $\hat h_4$ &
      $\hat h_4 -\alpha_{14} \hat h_5$ & $\hat h_4$ \\
      $\hat h_5$ & $\hat h_5$ & $\hat h_5$ & $\hat h_5$ & $\hat h_5$ &
      $\hat h_5 \exp{\left(-2\alpha_{13}\right)}$ & $\hat h_5$ &
      $\hat h_5 -\alpha_{15} \hat h_4$ \\
      $\hat h_6$ & $\hat h_6 +2\alpha_9 \hat h_{12} +4\alpha_9^2 \hat h_9$ &
      $\hat h_6$ &
      $\hat h_6 +2\alpha_{11} \hat h_{14} + \alpha_{11}^2 \hat h_{10}$ &
      $\hat h_6 \exp{\left(4\alpha_{12}\right)}$ & $\hat h_6$ & $\hat h_6$ &
      $\hat h_6 +2\alpha_{15} \hat h_8 +\alpha_{15}^2 \hat h_7$ \\
      $\hat h_7$ & $\hat h_7$ &
      $\hat h_7 +2\alpha_{10} \hat h_{13} +4\alpha_{10}^2 \hat h_{10}$ &
      $\hat h_7 +2\alpha_{11} \hat h_{15} + \alpha_{11}^2 \hat h_9$ &
      $\hat h_7$ & $\hat h_7 \exp{\left(4\alpha_{13}\right)}$ &
      $\hat h_7 +2\alpha_{14} \hat h_8 +\alpha_{14}^2 \hat h_6$ & $\hat h_7$ \\
      $\hat h_8$ & $\hat h_8 +2\alpha_9 \hat h_{15}$ &
      $\hat h_8 +2\alpha_{10} \hat h_{14}$ & $\hat h_8 +\alpha_{11}
      \frac{\hat h_{12}+\hat h_{13}}{2}+\alpha_{11}^2\hat h_{11}$ &
      $\hat h_8 \exp{\left(2\alpha_{12}\right)}$ &
      $\hat h_8 \exp{\left(2\alpha_{13}\right)}$ &
      $\hat h_8 +\alpha_{14} \hat h_6$ & $\hat h_8 +\alpha_{15} \hat h_7$ \\
      $\hat h_9$ & $\hat h_9$ & $\hat h_9$ & $\hat h_9$ & 
      $\hat h_9 \exp{\left(-4\alpha_{12}\right)}$ & $\hat h_9$ &
      $\hat h_9 -2\alpha_{14} \hat h_{11} +\alpha_{14}^2 \hat h_{10}$ &
      $\hat h_9$ \\
      $\hat h_{10}$ & $\hat h_{10}$ & $\hat h_{10}$ & $\hat h_{10}$ &
      $\hat h_{10}$ & $\hat h_{10} \exp{\left(-4\alpha_{13}\right)}$ &
      $\hat h_{10}$ &
      $\hat h_{10} -2\alpha_{15} \hat h_{11} +\alpha_{15}^2 \hat h_9$ \\
      $\hat h_{11}$ & $\hat h_{11}$ & $\hat h_{11}$ & $\hat h_{11}$ &
      $\hat h_{11} \exp{\left(-2\alpha_{12}\right)}$ &
      $\hat h_{11} \exp{\left(-2\alpha_{13}\right)}$ &
      $\hat h_{11} -\alpha_{14} \hat h_{10}$ &
      $\hat h_{11} -\alpha_{15} \hat h_9$ \\
      $\hat h_{12}$ & $\hat h_{12} +4\alpha_9 \hat h_9$ & $\hat h_{12}$ &
      $\hat h_{12} +2\alpha_{11} \hat h_{11}$ & $\hat h_{12}$ & $\hat h_{12}$ &
      $\hat h_{12} -2\alpha_{14} \hat h_{14}$ &
      $\hat h_{12} +2\alpha_{15} \hat h_{15}$ \\
      $\hat h_{13}$ & $\hat h_{13}$ & $\hat h_{13} +4\alpha_{10} \hat h_{10}$ &
      $\hat h_{13} +2\alpha_{11} \hat h_{11}$ & $\hat h_{13}$ & $\hat h_{13}$ &
      $\hat h_{13} +2\alpha_{14} \hat h_{14}$ &
      $\hat h_{13} -2\alpha_{15} \hat h_{15}$ \\
      $\hat h_{14}$ & $\hat h_{14} +2\alpha_9 \hat h_{11}$ & $\hat h_{14}$ &
      $\hat h_{14} +\alpha_{11} \hat h_{10}$ &
      $\hat h_{14} \exp{\left(2\alpha_{12}\right)}$ &
      $\hat h_{14} \exp{\left(-2\alpha_{13}\right)}$ & $\hat h_{14}$ &
      $\hat h_{14} -\alpha_{15} \frac{\hat h_{12} -\hat h_{13}}{2}
      -\alpha_{15}^2 \hat h_{15}$ \\
      $\hat h_{15}$ & $\hat h_{15}$ & $\hat h_{15} +2\alpha_{10} \hat h_{11}$ &
      $\hat h_{15} +\alpha_{11} \hat h_9$ &
      $\hat h_{15} \exp{\left(-2\alpha_{12}\right)}$ &
      $\hat h_{15} \exp{\left(2\alpha_{13}\right)}$ &
      $\hat h_{15} +\alpha_{14} \frac{\hat h_{12} -\hat h_{13}}{2}
      -\alpha_{14}^2 \hat h_{14}$ & $\hat h_{15}$ \\
      \hline
      $\hat p_t$ & $\hat p_t +\dot \alpha_{9} \hat h_{9}$ &
      $\hat p_t +\dot \alpha_{10} \hat h_{10}$ &
      $\hat p_t +\dot \alpha_{11} \hat h_{11}$ &
      $\hat p_t +\dot \alpha_{12} \hat h_{12}$ &
      $\hat p_t +\dot \alpha_{13} \hat h_{13}$ &
      $\hat p_t +\dot \alpha_{14} \hat h_{14}$ &
      $\hat p_t +\dot \alpha_{15} \hat h_{15}$
    \end{tabular}}
  \end{center}
\end{table}

Now we can start to build the evolution operator by identifying
the elements of the unitary group generated by $\mathcal{L}_{15}$.
The transformations produced by the 15 generators of
$\mathcal{L}_{15}$ are given by
\begin{align}
\hat U_1 &=\exp\left(i\alpha_1\hat 1/\hbar\right),&
\hat U_2 &=\exp\left(i\alpha_2\hat x/\hbar\right),&
\hat U_3 &=\exp\left(i\alpha_3\hat y/\hbar\right),\nonumber\\
\hat U_4 &=\exp\left(i\alpha_4\hat p_x/\hbar\right),&
\hat U_5 &=\exp\left(i\alpha_5\hat p_y/\hbar\right),&
\hat U_6 &=\exp\left(i\alpha_6\hat x^2/\hbar\right),\nonumber\\
\hat U_7 &=\exp\left(i\alpha_7\hat y^2/\hbar\right),&
\hat U_8 &=\exp\left(i\alpha_8\hat x\hat y/\hbar\right),&
\hat U_9 &=\exp\left(i\alpha_9\hat p_x^2/\hbar\right),\\
\hat U_{10} &=\exp\left(i\alpha_{10}\hat p_y^2/\hbar\right),&
\hat U_{11} &=\exp\left(i\alpha_{11}\hat p_x\hat p_y/\hbar\right),&
\hat U_{12} &=\exp\left[i\alpha_{12}\left(\hat x\hat p_x+\hat p_x\hat x\right)/\hbar\right],\nonumber\\
\hat U_{13} &=\exp\left[i\alpha_{13}\left(\hat y\hat p_y+\hat p_y\hat y\right)/\hbar\right],&
\hat U_{14} &=\exp\left(i\alpha_{14}\hat x\hat p_y/\hbar\right),&
\hat U_{15} &=\exp\left(i\alpha_{15}\hat y\hat p_x/\hbar\right).\nonumber
\end{align}
The first five transformations shift the energy, position and momentum operators
by the time-dependent functions $\alpha_1$ to $\alpha_5$ . We
prove later on that these parameters are related with the classical
action $S$, position $ x$, $ y$ and momentum $- p_x$, $-p_y$.
Whereas $\hat U_6$ and $\hat U_7$ shift the momentum operator
by $-\alpha_6\hat x$ and $-\alpha_7\hat y$, $\hat U_9$
and $\hat U_{10}$ shift the position operator by $-\alpha_9\hat p_x$
and $\alpha_{10}\hat p_y$ .
The dilatations $\hat U_{12}$ and $\hat U_{13}$ preserve the commutation relations
between the transformed position and momentum operators by expanding
the position operators
$\hat x$ and $\hat y$ by the factors
$\exp(2\alpha_{12})$ and $\exp(2\alpha_{13})$, respectively, while
contracting the momentum operators by the inverse
factors $\exp(-2\alpha_{12})$ and
$\exp(-2\alpha_{13})$, respectively.
In principle it is possible to compute the 240 transformation rules
summarized in Tables \ref{transf:1-8} and \ref{transf:9-15}, however in order to
obtain them all, only a few are needed.
Eq. (\ref{enerrules:gen}) yields the first 15
transformation rules. Eq. (\ref{re:cuan}) allows to derive
the transformation rules
for all the remaining transformations except the dilatations $\hat U_{12}$ and $\hat U_{13}$.
For example
\begin{equation}
\hat U_9\hat x\hat U_9^\dagger=\hat x+\hat U_9\left[\hat x,\hat U_9^\dagger\right]
=\hat x+\hat U_9\left[\hat x,\hat p_x\right]
\frac{\partial \hat U_9^\dagger}{\partial \hat p_x}=\hat x+2\alpha_9\hat p_x.
\end{equation}
On the other hand, 
the action of the dilatations is better
calculated by taking the derivative with respect to the transformation
parameter. Then, for the dilatation $\hat U_{12}$ we have
\begin{equation}
\frac{\partial }{\partial \alpha_{12}}\hat U_{12}\hat x\hat U_{12}^\dagger=
\frac{i}{\hbar}\hat U_{12}\left[\hat x,\hat x\hat p_x
+\hat p_x\hat x\right]\hat U_{12}^\dagger
=2\hat U_{12}\hat x\hat U_{12}^\dagger,
\end{equation} 
hence, integrating we get
\begin{equation}
\hat U_{12}\hat x\hat U_{12}^\dagger=\exp(2\alpha_{12})\hat x,
\end{equation}
given that $\hat U_{12}\hat x\hat U_{12}^{\dagger}=\hat x$ for $\alpha_{12}=0$.
The action of dilatations on the momentum operators is
obtained in a similar way by deriving
$\hat U_{12}\hat p_x\hat U_{12}^{\dagger}$
with respect to the transformation parameter $\alpha_{12}$.
The $\mathcal{M}_i$ matrices corresponding to these transformation rules
are presented for reference in the
supplemental material at [URL will be inserted by AIP].

Let us now reduce the Floquet operator of the generalized two-dimensional
quadratic Hamiltonian in Eq. (\ref{ham:gao})
by means of these unitary transformations. We thus calculate the transformed
Floquet operator by applying the unitary transformation
$\hat U=\hat U_{15}\hat U_{14}\hat U_{13}\hat U_{12}\hat U_{11}
\hat U_9\hat U_8\hat U_7\hat U_6\hat U_5\hat U_4\hat U_3\hat U_2\hat U_1$
stepwisely.
Proceeding in this way through the 15 transformations we obtain 
\begin{multline}
\hat U\hat {\mathcal{H}}\hat U^\dagger =
\hat U\left(\hat H-\hat p_t\right)\hat U^\dagger
=u_1\hat h_1+u_2\hat h_2+u_3\hat h_3+u_4\hat h_4+u_5\hat h_5+u_6\hat h_6
+u_7\hat h_7\\+u_8\hat h_8+u_9\hat h_9+u_{10}\hat h_{10}+u_{11}\hat h_{11}
+u_{12}\hat h_{12}+u_{13}\hat h_{13}+u_{14}\hat h_{14}+u_{15}\hat h_{15}
-\hat p_t,
\label{fho:reducedfloquet}
\end{multline}
where the explicit form of the $u$ coefficients as functions of
 $a$, $\alpha$ and $\dot\alpha$ is to involved to
be presented here (see supplemental material at [URL will be inserted by AIP]
for the explicit form of the $u$ coefficients).
However, the upshot of the method presented
in Section \ref{liealap} is that by calculating the matrix $\nu$ through
Eq. (\ref{nudef}) and using (\ref{uiformtwo}) it is possible to express the $u$
coefficients as compact functions of
$\mathcal E$ and $\alpha$ in the following form
\begin{eqnarray}
u_1 &=&\mathcal{E}_1+ \alpha _4 \mathcal{E}_2+\alpha _5 \mathcal{E}_3,\label{u:eq1}
 \\
u_2 &=& e^{2 \alpha _{12}} \mathcal{E}_2
   +e^{2 \alpha _{13}} \alpha _{14}\mathcal{E}_3
   -\left(2 e^{2 \alpha _{12}} \alpha _6+e^{2\alpha _{13}} \alpha _8 \alpha _{14}\right) \mathcal{E}_4
   -\left(e^{2 \alpha _{12}} \alpha _8+2 e^{2\alpha _{13}} \alpha _7 \alpha _{14}\right) \mathcal{E}_5,
   \\
u_3 &=& e^{2 \alpha _{12}} \alpha _{15} \mathcal{E}_2
   +e^{2 \alpha_{13}} \left(\alpha _{14} \alpha _{15}+1\right)\mathcal{E}_3
   -\left[e^{2 \alpha _{12}} \alpha _8
   \alpha _{15}+2 e^{2 \alpha _{13}} \alpha _7
   \left(\alpha _{14} \alpha _{15}+1\right)\right]\mathcal{E}_5\nonumber\\
  && -\left[2 e^{2 \alpha _{12}} \alpha _6
   \alpha _{15}+e^{2 \alpha _{13}} \alpha _8
   \left(\alpha_{14} \alpha _{15}+1\right)\right] \mathcal{E}_4, \\
u_4 &=& \left[2 e^{-2 \alpha _{12}} \alpha _9 \left(\alpha _{14} \alpha_{15}+1\right)
             -e^{-2 \alpha _{13}} \alpha _{11} \alpha_{15}\right] \mathcal{E}_2
 \nonumber \\
   &&+\left[e^{-2 \alpha _{12}} \alpha_{11}
   \left(\alpha _{14} \alpha _{15}+1\right)-2 
   e^{-2 \alpha_{13}} \alpha _{10} \alpha _{15}\right]\mathcal{E}_3
\nonumber \\
    &&+\left[2 e^{-2 \alpha _{13}} \left(\alpha _8
   \alpha _{10}+\alpha _6 \alpha _{11}\right) \alpha _{15}
   -e^{-2\alpha _{12}} \left(4 \alpha _6 \alpha _9+\alpha _8 \alpha_{11}-1\right)
    \left(\alpha _{14} \alpha _{15}+1\right)\right] \mathcal{E}_4
\nonumber \\
   &&+\left[e^{-2 \alpha _{13}} \left(4 \alpha _7
   \alpha _{10}+\alpha _8 \alpha _{11}-1\right) \alpha _{15}-2
   e^{-2 \alpha _{12}} \left(\alpha _8 \alpha _9+\alpha _7 \alpha_{11}\right)
    \left(\alpha _{14} \alpha _{15}+1\right)\right]\mathcal{E}_5,\\
u_5 &=& \left(e^{-2 \alpha _{13}} \alpha _{11}-2 e^{-2 \alpha _{12}}
   \alpha _9 \alpha _{14}\right) \mathcal{E}_2
   +\left(2 e^{-2\alpha _{13}} \alpha _{10}-e^{-2 \alpha _{12}} \alpha _{11}
   \alpha _{14}\right) \mathcal{E}_3
\nonumber \\
&&   +\left[e^{-2 \alpha _{12}} \left(4 \alpha _6
   \alpha _9+\alpha _8 \alpha _{11}-1\right) \alpha _{14}-2 e^{-2
   \alpha _{13}} \left(\alpha _8 \alpha _{10}
   +\alpha _6 \alpha_{11}\right)\right] \mathcal{E}_4
\nonumber \\
&&+\left[2 e^{-2 \alpha _{12}}
   \left(\alpha _8 \alpha _9+\alpha _7 \alpha _{11}\right) \alpha _{14}
   -e^{-2 \alpha _{13}} \left(4 \alpha _7 \alpha_{10}
   +\alpha _8 \alpha _{11}-1\right)\right]\mathcal{E}_5,\\
u_6 &=& e^{4 \alpha _{12}} \alpha _{15}^2 \mathcal{E}_6
      +e^{4 \alpha _{13}} \alpha _{14}^2 \mathcal{E}_7
     +e^{2 \left(\alpha_{12}+\alpha _{13}\right)} \alpha _{14} \mathcal{E}_8,\\
u_7 &=& e^{4 \alpha _{12}} \alpha _{15}^2 \mathcal{E}_6
 +e^{4 \alpha _{13}}\left(\alpha _{14} \alpha _{15}+1\right)^2 \mathcal{E}_7
    +e^{2 \left(\alpha_{12}+\alpha _{13}\right)} 
    \left(\alpha _{14} \alpha_{15}+1\right) \alpha _{15} \mathcal{E}_8
    ,\\
u_8 &=& 2 e^{4 \alpha _{12}} \alpha _{15} \mathcal{E}_6+2 e^{4 \alpha
   _{13}} \alpha _{14} \left(\alpha _{14} \alpha _{15}+1\right)
   \mathcal{E}_7+e^{2 \left(\alpha _{12}+\alpha _{13}\right)}
   \left(2 \alpha _{14} \alpha _{15}+1\right) \mathcal{E}_8,\\
u_9 &=& 
    e^{-4 \left(\alpha _{12}+\alpha _{13}\right)}
   \left[e^{2 \alpha _{12}} \alpha _{11} \alpha _{15}-2 e^{2\alpha _{13}} \alpha _9 
   \left(\alpha _{14} \alpha_{15}+1\right)\right]^2 \mathcal{E}_6
   \nonumber \\
   &&+e^{-4 \left(\alpha_{12}+\alpha _{13}\right)}
   \left[e^{2 \alpha _{13}} \alpha_{11} \left(\alpha _{14} \alpha _{15}+1\right)
   -2 e^{2 \alpha_{12}} \alpha _{10} \alpha _{15}\right]^2\mathcal{E}_7
   \nonumber \\
   &&+e^{-4 \left(\alpha _{12}+\alpha _{13}\right)}\left[2 e^{2 \alpha _{13}} \alpha _9
    \left(\alpha _{14} \alpha_{15}+1\right)-e^{2 \alpha _{12}} \alpha _{11} \alpha_{15}\right]
 \nonumber \\
    &&\times
    \left[e^{2 \alpha _{13}} \alpha _{11}\left(\alpha _{14} \alpha _{15}+1\right)
    -2 e^{2 \alpha _{12}}\alpha _{10} \alpha _{15}\right] \mathcal{E}_8
 \nonumber \\
   &&+e^{-4\alpha _{12}} \left(\alpha _{14} \alpha _{15}+1\right)^2\mathcal{E}_9 
   +e^{-4 \alpha _{13}} \alpha _{15}^2 \mathcal{E}_{10}
   -e^{-2\left(\alpha _{12}+\alpha _{13}\right)} \left(\alpha _{14}\alpha _{15}+1\right)
    \alpha _{15} \mathcal{E}_{11},\\
u_{10} &=&
   e^{-4 \left(\alpha _{12}+\alpha_{13}\right)}
      \left(e^{2 \alpha _{12}} \alpha _{11}-2e^{2 \alpha _{13}}
       \alpha _9 \alpha _{14}\right)^2\mathcal{E}_6
       +e^{-4 \left(\alpha _{12}+\alpha_{13}\right)}
        \left(e^{2 \alpha _{13}} \alpha _{11}
        \alpha _{14}-2 e^{2 \alpha _{12}} \alpha_{10}\right)^2 \mathcal{E}_7
\nonumber \\
        &&+e^{-4 \left(\alpha _{12}+\alpha_{13}\right)}
         \left(2 e^{2 \alpha _{13}} \alpha _9
   \alpha _{14}-e^{2 \alpha _{12}} \alpha _{11}\right)
   \left(e^{2 \alpha _{13}} \alpha _{11} \alpha _{14}-2
   e^{2 \alpha _{12}} \alpha _{10}\right) \mathcal{E}_8
   +e^{-4 \alpha _{12}} \alpha _{14}^2 \mathcal{E}_9
\nonumber \\
    &&+e^{-4 \alpha _{13}}\mathcal{E}_{10}
   -e^{-2\left(\alpha _{12}+\alpha _{13}\right)} \alpha _{14}\mathcal{E}_{11} ,\\
u_{11} &=&- \left[ 2 e^{-4 \left(\alpha _{12}+\alpha _{13}\right)} \left(e^{2 \alpha _{12}} \alpha
   _{11}-2 e^{2 \alpha _{13}} \alpha _9 \alpha _{14}\right) \left(e^{2 \alpha
   _{12}} \alpha _{11} \alpha _{15}-2 e^{2 \alpha _{13}} \alpha _9 \left(\alpha
   _{14} \alpha _{15}+1\right)\right) \right]\mathcal{E}_6
\nonumber \\           
   &&-2 e^{-4 \left(\alpha _{12}+\alpha_{13}\right)}
  \left(e^{2 \alpha _{13}} \alpha _{11} \alpha_{14}-2 e^{2 \alpha _{12}} \alpha _{10}\right)
    \left[e^{2\alpha _{13}} \alpha _{11} \left(\alpha _{14} \alpha_{15}+1\right)
  -2 e^{2 \alpha _{12}} \alpha _{10} \alpha_{15}\right]\mathcal{E}_7
\nonumber \\
    &&+\left[e^{-2 \left(\alpha _{12}+\alpha _{13}\right)}
    \left(\alpha _{11}^2+4 \alpha _9 \alpha _{10}\right)
    \left(2\alpha _{14} \alpha _{15}+1\right)-4 e^{-4 \alpha _{13}}
   \alpha _{10} \alpha_{11} \alpha _{15}\right.
 \nonumber \\
   &&\left.-4 e^{-4 \alpha _{12}} \alpha _9 \alpha_{11} \alpha _{14} 
   \left(\alpha _{14} \alpha_{15}+1\right)
    \right] \mathcal{E}_8
   -2 e^{-4\alpha _{12}} \alpha _{14} \left(\alpha _{14} \alpha_{15}+1\right) \mathcal{E}_9
\nonumber \\
   && -2 e^{-4 \alpha _{13}} \alpha _{15} \mathcal{E}_{10}
   +e^{-2 \left(\alpha _{12}+\alpha_{13}\right)} \left(2 \alpha _{14} \alpha _{15}+1\right)\mathcal{E}_{11}
\end{eqnarray}
\begin{eqnarray}
u_{12} &=&
  \left[2 \alpha _9 \left(\alpha_{14} \alpha _{15}+1\right)-e^{2 \alpha _{12}-2 \alpha _{13}}
   \alpha _{11} \alpha _{15}\right] \mathcal{E}_6
   +\alpha _{14}\left[e^{2 \alpha _{13}-2 \alpha _{12}} \alpha _{11}
   \left(\alpha _{14} \alpha _{15}+1\right)-2 \alpha _{10} \alpha_{15}\right] \mathcal{E}_7
\nonumber \\
   &&+\left[\frac{\alpha _{11}}{2}-e^{2\alpha _{12}-2 \alpha _{13}} \alpha _{10} \alpha _{15}
   +e^{2\alpha _{13}-2 \alpha _{12}} \alpha _9 \alpha _{14}
    \left(\alpha _{14} \alpha _{15}+1\right)\right]\mathcal{E}_8
  +\left(\alpha _{14} \alpha_{15}+1\right) \mathcal{E}_{12}
\nonumber \\
  &&  -\frac{1}{2} \alpha _{15} \mathcal{E}_{14}-\alpha _{14} \alpha_{15} \mathcal{E}_{13},\\
u_{13} &=&
  \left(e^{2 \alpha _{12}-2 \alpha _{13}}\alpha _{11}-2 \alpha _9 \alpha _{14}\right)
   \alpha _{15}\mathcal{E}_6
   +e^{-2 \alpha _{12}} \left(2 e^{2 \alpha_{12}} \alpha _{10}
   -e^{2 \alpha _{13}} \alpha _{11} \alpha_{14}\right)
    \left(\alpha _{14} \alpha _{15}+1\right)\mathcal{E}_7
 \nonumber \\
   &&+\left[\frac{\alpha _{11}}{2}+e^{2 \alpha _{12}-2\alpha _{13}} \alpha _{10} \alpha _{15}
    -e^{2 \alpha _{13}-2\alpha _{12}}
     \alpha _9 \alpha _{14} \left(\alpha _{14} \alpha_{15}+1\right)\right] \mathcal{E}_8
      -\alpha _{14} \alpha _{15}\mathcal{E}_{12}
  \nonumber \\
&& +\left(\alpha _{14} \alpha _{15}+1\right)\mathcal{E}_{13}
    +\frac{\alpha _{15}}{2} \mathcal{E}_{14},\\
u_{14} &=&
 \left(2 e^{2 \alpha _{12}-2 \alpha _{13}}\alpha _{11}-4 \alpha _9 \alpha _{14}\right)\mathcal{E}_6
    +\left(4 \alpha_{10} \alpha _{14}-2 e^{2 \alpha _{13}-2 \alpha _{12}} 
    \alpha_{11} \alpha _{14}^2\right) \mathcal{E}_7
 \nonumber \\
  &&+e^{-2 \left(\alpha _{12}+\alpha _{13}\right)}
   \left(2 e^{4 \alpha _{12}} \alpha _{10}-2 e^{4 \alpha _{13}}
   \alpha _9 \alpha _{14}^2\right) \mathcal{E}_8
  -2 \alpha _{14} \mathcal{E}_{12}
  +2 \alpha _{14}\mathcal{E}_{13}
  +\mathcal{E}_{14},\\
u_{15} &=& 
   2 \left[2\alpha _9 \left(\alpha _{14} \alpha_{15}+1\right)
   -e^{2 \alpha _{12}-2 \alpha _{13}}\alpha _{11} \alpha _{15}\right]
   \alpha _{15}\mathcal{E}_6\nonumber \\
   &&+2 \left(\alpha _{14} \alpha_{15}+1\right) \left[e^{2 \alpha _{13}-2 \alpha _{12}}
   \alpha _{11} \left(\alpha _{14} \alpha_{15}+1\right)
   -2 \alpha _{10} \alpha _{15}\right]\mathcal{E}_7\nonumber \\
    &&+2 e^{-2 \left(\alpha _{12}+\alpha_{13}\right)} \left[e^{4 \alpha _{13}} \alpha _9
   \left(\alpha _{14} \alpha _{15}+1\right)^2-e^{4\alpha _{12}}
   \alpha _{10} \alpha _{15}^2\right]\mathcal{E}_8
   +2\left(\alpha _{14} \alpha _{15}+1\right) \alpha _{15}\mathcal{E}_{12}\nonumber\\
   &&-2 \left(\alpha _{14} \alpha_{15}+1\right) \alpha _{15} \mathcal{E}_{13}
   -\alpha _{15}^2\mathcal{E}_{14}
   +\mathcal{E}_{15}.
\end{eqnarray}
The $\nu$ matrix can be obtained from the previous equations
by using the right-hand side of Eq. (\ref{nudef}).

For the particular transformation ordering used here
$\det \nu=1$, thus upon calculating the inverse of $\nu$,
the explicit form of the equations $\mathcal E$ 
if readily obtained by using Eq. (\ref{equations})
\begin{eqnarray}
\mathcal{E}_1 &=& a_9 \alpha _2^2-a_4 \alpha _2+a_{11} \alpha _3 \alpha _2+a_{10}
   \alpha _3^2-a_6 \alpha _4^2-a_7 \alpha _5^2-a_5 \alpha _3
   -a_8\alpha _4 \alpha _5+a_1-\dot{\alpha }_1,\label{difeq:01}\\
\mathcal{E}_2 &=&-2 a_{12} \alpha _2-a_{14} \alpha _3
  +2 a_6\alpha _4+a_8 \alpha _5+a_2-\dot{\alpha }_2 ,\label{difeq:02}\\
\mathcal{E}_3 &=& -a_{15} \alpha _2-2 a_{13} \alpha _3
  +a_8 \alpha _4+2 a_7 \alpha _5+a_3-\dot{\alpha }_3,\label{difeq:03}\\
\mathcal{E}_4 &=& -2 a_9 \alpha _2-a_{11} \alpha _3+2 a_{12} \alpha _4+a_{15} \alpha 
_5+a_4-\dot{\alpha }_4,\\
\mathcal{E}_5 &=& -a_{11} \alpha _2-2 a_{10} \alpha _3+a_{14} \alpha _4
  +2 a_{13} \alpha _5+a_5-\dot{\alpha }_5,\label{difeq:05}\\
\mathcal{E}_6 &=& 4 a_9 \alpha _6^2-4 a_{12} \alpha _6
  +2 a_{11} \alpha _8 \alpha_6+a_{10} \alpha _8^2-a_{14} \alpha _8+a_6-\dot{\alpha }_6,\label{difeq:06}\\
\mathcal{E}_7 &=& 4 a_{10} \alpha _7^2-4 a_{13} \alpha _7
  +2 a_{11} \alpha _8 \alpha_7+a_9 \alpha _8^2-a_{15} \alpha _8+a_7-\dot{\alpha }_7,\\
\mathcal{E}_8 &=& - 2 a_{14}\alpha _7 -2 a_{15} \alpha_6-2 a_{12} \alpha _8-2 a_{13} \alpha _8
   +4a_9 \alpha _6 \alpha _8+4 a_{10} \alpha _7 \alpha _8
\nonumber \\
   &&+a_{11}\left(\alpha _8^2+4 \alpha _6 \alpha _7\right)+a_8-\dot{\alpha}_8,\label{difeq:08}\\
\mathcal{E}_9 &=& 4 a_{12} \alpha _9+a_{15} \alpha _{11}
   -2 a_{11} \left(\alpha _8\alpha _9+\alpha _7 \alpha _{11}\right)
   +a_9 \left(1-8 \alpha _6\alpha _9-2 \alpha _8 \alpha _{11}\right)-\dot{\alpha }_9,\label{difeq:09}\\
\mathcal{E}_{10} &=& 4 a_{13} \alpha _{10}+a_{14} \alpha _{11}
   -2 a_{11} \left(\alpha_8 \alpha _{10}+\alpha _6 \alpha _{11}\right)
   +a_{10} \left(1-8\alpha _7 \alpha _{10}-2 \alpha _8 \alpha_{11}\right)-\dot{\alpha }_{10},\\
\mathcal{E}_{11} &=& 2 a_{14} \alpha _9
  +2 a_{15}\alpha _{10}+2 a_{12} \alpha _{11}+2 a_{13} \alpha _{11}
\nonumber \\ 
  &&-a_9 \left[4 \alpha _8 \alpha _{10}+4 \alpha _6 \alpha _{11}\right]
  -a_{10}\left[4 \alpha _8 \alpha _9 + 4 \alpha_7 \alpha _{11}\right]
\nonumber \\
  &&+a_{11} \left[1-4 \alpha _6 \alpha _9
  -4 \alpha _7 \alpha_{10}-2 \alpha _8 \alpha _{11}\right]-\dot{\alpha }_{11},\label{difeq:11}\\
\mathcal{E}_{12} &=& \frac{1}{2} e^{2 \alpha _{13}-2 \alpha _{12}} a_{15} \alpha_{14}
    -a_{11} \left(\frac{\alpha _8}{2}+e^{2 \alpha _{13}-2\alpha _{12}} \alpha _7 \alpha _{14}\right)
\nonumber \\
   && -a_9 \left(2\alpha _6+e^{2 \alpha _{13}-2 \alpha _{12}} \alpha _8 \alpha_{14}\right)
   +a_{12}-\dot{\alpha }_{12},\label{difeq:12}\\
\mathcal{E}_{13} &=& -2 a_{10} \alpha _7
   -\frac{1}{2} e^{2 \alpha _{13}-2 \alpha _{12}}a_{15} \alpha _{14}
   +e^{2 \alpha _{13}-2 \alpha _{12}} a_9\alpha_8 \alpha _{14}
\nonumber \\
   &&+a_{11} \left(e^{2 \alpha _{13}-2 \alpha_{12}} \alpha _7 \alpha _{14}
   -\frac{\alpha_8}{2}\right)+a_{13}-\dot{\alpha }_{13},\\
\mathcal{E}_{14} &=& e^{2 \alpha _{13}-2 \alpha _{12}} a_{15} \alpha _{14}^2
  -2 e^{2\alpha _{13}-2 \alpha _{12}} a_9 \alpha _8 \alpha _{14}^2
  +e^{2\alpha _{12}-2 \alpha _{13}} a_{14}
  -2 e^{2 \alpha _{12}-2\alpha _{13}} a_{10} \alpha _8
\nonumber \\
  &&-2 e^{-2 \left(\alpha_{12}+\alpha _{13}\right)} a_{11} \left(e^{4 \alpha _{13}}\alpha _7 \alpha _{14}^2
  +e^{4 \alpha _{12}} \alpha_6\right)-\dot{\alpha }_{14},\label{difeq:14}\\
\mathcal{E}_{15} &=& e^{2 \alpha _{13}-2 \alpha _{12}} a_{15}
  -2 e^{2 \alpha _{13}-2\alpha _{12}} a_{11} \alpha_7
  -2 e^{2 \alpha _{13}-2 \alpha_{12}} a_9 \alpha _8-\dot{\alpha }_{15}.\label{difeq:15}
\end{eqnarray}
Here it is important to stress that different transformation orderings
yield different $\det \nu$ values. For this presentation we have chosen
a transformation ordering that yields $\det \nu=1$ which considerably
simplifies the previous expressions.

Now, since the elements of $u$ are expressible as a linear combination
of the differential equations $\mathcal E$, and
since each one of them satisfies Eq. (\ref{ode:sys}) with $\det\nu=1\ne 0$,
$\mathcal E$ vanishes identically, and the final condition
becomes merely Eq. (\ref{equations}) or alternatively
$\mathcal{E}_1=\mathcal{E}_2=\dots \mathcal{E}_{15}=0$.
Additionally we must impose the initial condition that the
transformation parameters vanish at $t=0$
\begin{equation}
\alpha_i(0)=0, \,\,\,\,\,\,\,\ i=1,2,\dots 15,
\label{ini:cond}
\end{equation}
in order to ensure that the evolution operator equals the identity at
$t=0$,  namely $\hat {\mathcal{U}}(0)=\hat 1$ and therefore
$\hat U_1(0)=\hat U_2(0)=\dots \hat U_{15}(0)=\hat 1$.

The solution to the system of ordinary differential equations
(\ref{difeq:01})-(\ref{difeq:15}), together with the initial conditions (\ref{ini:cond}),
yields the explicit form of the set of transformation parameters
as functions of time.

The first five differential equations (\ref{difeq:01})-(\ref{difeq:05})
may be solved independently
for the transformation parameters $\alpha_1$, $\alpha_2$, $\alpha_3$,
$\alpha_4$ and $\alpha_5$.
This is a direct consequence of the fact that $\hat h_1$, $\hat h_2$,
$\hat h_3$, $\hat h_4$ and $\hat h_5$
form a sub-algebra of $\mathcal{L}_{15}$
as can be verified in Table \ref{ta:commutators}.
There is a close analogy between these five equations and the
equations of motion of the classical version of (\ref{gfho}).
Replacing $\mathcal E_1$, $\mathcal E_2$ and $\mathcal E_3$ from
Eqs. (\ref{difeq:01}), (\ref{difeq:02}) and (\ref{difeq:03})
into $u_1$ in Eq. (\ref{u:eq1})
we get
\begin{multline}
u_1 = a_9 \alpha _2^2-a_4 \alpha _2+a_{11} \alpha _3 \alpha _2-2 a_{12}
   \alpha _4 \alpha _2-a_{15} \alpha _5 \alpha _2+a_{10} \alpha_3^2
   +a_6 \alpha _4^2+a_7 \alpha _5^2-a_5 \alpha _3\\
   +a_2 \alpha_4-a_{14} \alpha _3 \alpha _4
   +a_3 \alpha _5-2 a_{13} \alpha _3\alpha _5+a_8 \alpha _4 \alpha _5
   +a_1-\alpha _4\dot{\alpha }_2-\alpha _5 \dot{\alpha }_3-\dot{\alpha }_1,
\end{multline}
Since $u_1$ must vanish in order to reduce the Floquet
operator, we may identify $\alpha_1$
with the classical action $S=\alpha_1$ and therefore, from the above equation
the classical Lagrangian $L=u_1+\dot{\alpha}_1=\dot S$ is given by
\begin{multline}
L= a_9 \alpha _2^2-a_4 \alpha _2+a_{11} \alpha _3 \alpha _2-2 a_{12}
   \alpha _4 \alpha _2-a_{15} \alpha _5 \alpha _2+a_{10} \alpha_3^2
   +a_6 \alpha _4^2+a_7 \alpha _5^2-a_5 \alpha _3\\
   +a_2 \alpha_4-a_{14} \alpha _3 \alpha _4
   +a_3 \alpha _5-2 a_{13} \alpha _3\alpha _5+a_8 \alpha _4 \alpha _5
   +a_1-\alpha _4\dot{\alpha }_2-\alpha _5 \dot{\alpha }_3.
\end{multline}
This analogy goes even further.
Indeed the Euler equations arising from this Lagrangian yield
\begin{eqnarray}
\frac{d}{dt}\frac{\partial L}{\partial \dot\alpha_2}
-\frac{\partial L}{\partial\alpha_2}
   &=& -2 a_9 \alpha _2-a_{11} \alpha _3+2 a_{12} \alpha _4+a_{15} \alpha_5
   +a_4-\dot{\alpha }_4=0 =\mathcal{E}_4,\label{fho:euler:eq1}
\\
  \frac{d}{dt}\frac{\partial L}{\partial \dot\alpha_3}-\frac{\partial L}{\partial\alpha_3}
    &=& -a_{11} \alpha _2-2 a_{10} \alpha _3+a_{14} \alpha _4
  +2 a_{13} \alpha _5+a_5-\dot{\alpha }_5 = \mathcal{E}_5,\label{fho:euler:eq2}
\\
    \frac{d}{dt}\frac{\partial L}{\partial \dot\alpha_4}
    -\frac{\partial L}{\partial\alpha_4} &=& 
    2 a_{12} \alpha _2+a_{14} \alpha _3
  -2 a_6\alpha _4-a_8 \alpha _5-a_2+\dot{\alpha }_2=-\mathcal{E}_2
    =0,
\label{fho:euler:eq3}\\
\frac{d}{dt}\frac{\partial L}{\partial \dot\alpha_5}
-\frac{\partial L}{\partial\alpha_5}
&=&
 a_{15} \alpha _2+2 a_{13} \alpha _3
  -a_8 \alpha _4-2 a_7 \alpha _5-a_3+\dot{\alpha }_3=-\mathcal{E}_3=0,
\label{fho:euler:eq4}
\end{eqnarray}
which precisely correspond to Eqs (\ref{difeq:02})-(\ref{difeq:05}).
In these equations it is clear that there is a correspondence between the
transformation parameters and the classical position and momentum. In particular,
$\alpha_2$, $\alpha_3$, $\alpha_4$ and $\alpha_5$ may be identified with the classical position
and momentum variables $-p_x$, $-p_y$, $x$ and $y$ respectively. 

The remaining transformation parameters $\alpha_6$ to $\alpha_{15}$ are obtained
from the solution of the system of ordinary differential equations (\ref{difeq:06})-(\ref{difeq:15})
and the initial conditions (\ref{ini:cond}).

With all the transformation parameters at hand we can write
the evolution operator as
\begin{eqnarray}
\hat{\mathcal{U}}=\hat U^\dagger
&=&\exp\left(-\frac{i}{\hbar}\alpha_1\hat h_1\right)
\exp\left(-\frac{i}{\hbar}\alpha_2\hat h_2\right)
\dots
\exp\left(-\frac{i}{\hbar}\alpha_{15}\hat h_{15}\right).\label{evopL6}
\end{eqnarray}
This equation together with the transformation rules in Tables
\ref{transf:1-8} and \ref{transf:9-15} allow us to compute
the evolution of any operator belonging to the $\mathcal{L}_{15}$ algebra
through Eqs. (\ref{gen:heispict}) and (\ref{gen:heispictfun}).
For example, 
the Heisenberg picture position and momentum operators
are obtained from Eq. (\ref{gen:heispict}) by acting
$\hat U$ on the Schr\"odinger picture
position and momentum operators and following the transformation rules.
For the most general case we have
\begin{eqnarray}
\hat x_H &=&\hat U\hat x\hat U^\dagger=\hat U\hat h_2\hat U^\dagger
    =\alpha _4 \hat{h}_1+e^{2 \alpha _{12}} \hat{h}_2
    +e^{2\alpha _{12}} \alpha _{15} \hat{h}_3\nonumber \\
    &&+\left[2 e^{-2\alpha _{12}} \alpha _9
    \left(\alpha _{14} \alpha_{15}+1\right)
    -e^{-2 \alpha _{13}} \alpha _{11} \alpha_{15}\right] \hat{h}_4
    +\left(e^{-2 \alpha _{13}}\alpha_{11}-2 e^{-2 \alpha_{12}} \alpha_9
     \alpha_{14}\right) \hat{h}_5\nonumber \\
  &=&\alpha _4
     +e^{2 \alpha _{12}} \hat{x}+e^{2 \alpha_{12}} \alpha _{15} \hat{y}
  +\left[2 e^{-2 \alpha _{12}} \alpha _9
   \left(\alpha _{14} \alpha _{15}+1\right)
   -e^{-2 \alpha_{13}} \alpha _{11} \alpha _{15}\right]\hat{p}_x\nonumber \\
  && +\left(e^{-2 \alpha _{13}} \alpha _{11}
   -2e^{-2 \alpha _{12}} \alpha _9 \alpha _{14}\right)\hat{p}_y,
    \label{fho:xh:1}\\
\hat y_H &=& \hat U\hat y\hat U^\dagger=\hat U\hat h_3\hat U^\dagger
   =\alpha _5 \hat{h}_1+e^{2 \alpha _{13}} \alpha _{14}\hat{h}_2
   +e^{2 \alpha _{13}}
   \left(\alpha _{14} \alpha_{15}+1\right) \hat{h}_3\nonumber \\
   &&+\left[e^{-2 \alpha _{12}}
   \alpha _{11} \left(\alpha _{14} \alpha_{15}+1\right)
   -2 e^{-2 \alpha _{13}} \alpha _{10}\alpha _{15}\right] \hat{h}_4
   +\left(2 e^{-2 \alpha_{13}} \alpha _{10}-e^{-2 \alpha _{12}} \alpha _{11}
   \alpha _{14}\right) \hat{h}_5\nonumber \\
   &=&\alpha _5
   +e^{2 \alpha _{13}} \alpha _{14}\hat{x}
   +e^{2 \alpha _{13}} \left(\alpha _{14} \alpha_{15}+1\right) \hat{y}
   +\left[e^{-2 \alpha _{12}} \alpha _{11}
   \left(\alpha _{14} \alpha _{15}+1\right)
   -2 e^{-2\alpha _{13}} \alpha _{10} \alpha _{15}\right]\hat{p}_x\nonumber \\
   &&+\left(2 e^{-2 \alpha _{13}} \alpha_{10}
   -e^{-2 \alpha _{12}} \alpha _{11} \alpha_{14}\right) \hat{p}_y,
\label{fho:yh:1}\\
\hat p_{xH} &=& \hat U\hat p_x\hat U^\dagger=\hat U\hat h_4\hat U^\dagger
=-\alpha _2 \hat{h}_1
 -\left(2 e^{2 \alpha _{12}} \alpha_6
 +e^{2 \alpha _{13}} \alpha _8 \alpha _{14}\right)\hat{h}_2\nonumber \\
 &&-\left[2 e^{2 \alpha _{12}} \alpha _6 \alpha_{15}
 +e^{2 \alpha _{13}} \alpha _8 \left(\alpha _{14}
   \alpha _{15}+1\right)\right] \hat{h}_3\nonumber \\
   &&+\left[2 e^{-2\alpha _{13}} \left(\alpha _8 \alpha _{10}
   +\alpha _6\alpha _{11}\right) \alpha _{15}
   -e^{-2 \alpha _{12}}\left(4 \alpha _6 \alpha _9+\alpha _8 \alpha_{11}-1\right)
    \left(\alpha _{14} \alpha_{15}+1\right)\right] \hat{h}_4\nonumber \\
    &&+\left[e^{-2 \alpha_{12}} \left(4 \alpha _6 \alpha _9
    +\alpha _8 \alpha_{11}-1\right) \alpha _{14}
    -2 e^{-2 \alpha _{13}}\left(\alpha _8 \alpha _{10}
    +\alpha _6 \alpha_{11}\right)\right] \hat{h}_5\nonumber\\
&=&-\alpha _2
    +\left(-2 e^{2 \alpha_{12}} \alpha _6
    -e^{2 \alpha _{13}} \alpha _8 \alpha_{14}\right) \hat{x}
    -\left[2 e^{2 \alpha _{12}}
    \alpha _6 \alpha _{15}+e^{2 \alpha _{13}} \alpha _8
   \left(\alpha _{14} \alpha _{15}+1\right)\right]\hat{y}    
    \nonumber \\
   &&+\left[2 e^{-2 \alpha _{13}} \left(\alpha _8\alpha _{10}
   +\alpha _6 \alpha _{11}\right) \alpha_{15}
   -e^{-2 \alpha _{12}} \left(4 \alpha _6 \alpha_9+\alpha _8 \alpha _{11}-1\right)
    \left(\alpha _{14}\alpha _{15}+1\right)\right]\hat{p}_x\nonumber \\
   && +\left[e^{-2\alpha _{12}}
    \left(4 \alpha _6 \alpha _9+\alpha _8\alpha _{11}-1\right) \alpha _{14}
    -2 e^{-2 \alpha_{13}} \left(\alpha _8 \alpha _{10}
    +\alpha _6 \alpha_{11}\right)\right] \hat{p}_y,
\label{fho:pxh:1}\\
\hat p_{yH} &=& \hat U\hat p_y\hat U^\dagger=\hat U\hat h_5\hat U^\dagger
=-\alpha _3 \hat{h}_1
 -\left(e^{2 \alpha _{12}} \alpha_8
 +2 e^{2 \alpha _{13}} \alpha _7 \alpha _{14}\right)\hat{h}_2\nonumber\\
 &&-\left[e^{2 \alpha _{12}} \alpha _8 \alpha_{15}
 +2 e^{2 \alpha _{13}}
 \alpha _7 \left(\alpha_{14} \alpha _{15}+1\right)\right]\hat{h}_3\nonumber\\
 &&+\left[e^{-2 \alpha _{13}} \left(4 \alpha _7\alpha _{10}
 +\alpha _8 \alpha _{11}-1\right) \alpha_{15}-2 e^{-2 \alpha _{12}}
 \left(\alpha _8 \alpha_9+\alpha _7 \alpha _{11}\right) \left(\alpha _{14}
 \alpha _{15}+1\right)\right] \hat{h}_4\nonumber \\
&& +\left[2 e^{-2\alpha _{12}} \left(\alpha _8 \alpha _9+\alpha _7
 \alpha _{11}\right) \alpha _{14}-e^{-2 \alpha _{13}}
 \left(4 \alpha _7 \alpha _{10}+\alpha _8 \alpha_{11}-1\right)\right] \hat{h}_5\nonumber\\
 &=&-\alpha _3
   -\left(e^{2 \alpha_{12}} \alpha _8
   +2 e^{2 \alpha _{13}} \alpha _7 \alpha_{14}\right) \hat{x}
   -\left[e^{2 \alpha _{12}} \alpha_8 \alpha _{15}+2 e^{2 \alpha _{13}} \alpha _7
   \left(\alpha _{14} \alpha _{15}+1\right)\right]\hat{y}\nonumber \\
 &&+\left[e^{-2 \alpha _{13}} \left(4 \alpha _7
   \alpha _{10}+\alpha _8 \alpha _{11}-1\right) \alpha_{15}
   -2 e^{-2 \alpha _{12}} \left(\alpha _8 \alpha_9+\alpha _7 \alpha _{11}\right)
   \left(\alpha _{14}\alpha _{15}+1\right)\right] \hat{p}_x\nonumber \\
   &&+\left[2 e^{-2\alpha _{12}} \left(\alpha _8 \alpha _9+\alpha _7
   \alpha _{11}\right) \alpha _{14}-e^{-2 \alpha _{13}}
   \left(4 \alpha _7 \alpha _{10}+\alpha _8 \alpha_{11}-1\right)\right] \hat{p}_y.\label{fho:pyh:1}
\end{eqnarray}

The propagator associated to the evolution operator (\ref{evopL6})
can be readily evaluated by separating the individual propagators
corresponding to each of the 15 unitary transformations
\begin{multline}
G(x,y,t; x^\prime, y^\prime,0)=
 \left\langle x,y\left\vert\hat{U}^\dag\left(t\right)\right\vert x^\prime,y^\prime \right\rangle\\
 =\int dx_1 dy_1\int dx_2 dy_2\dots\int dx_{n-1}dy_{n-1}
 \left\langle x,y \left\vert\hat{U}_1^\dag\left(t\right)\right\vert x_1, y_1 \right\rangle
 \left\langle x_1,y_1\left\vert\hat{U}_2^\dag\left(t\right)\right\vert x_2,y_2 \right\rangle\\
\times \dots\left\langle x_{n-2},y_{n-2}\left\vert\hat{U}_{n-1}^\dag\left(t\right)\right\vert 
 x_{n-1},y_{n-1}\right\rangle
 \left\langle x_{n-1},y_{n-1}\left\vert\hat{U}_{n}^\dag\left(t\right)\right\vert 
 x^\prime,y^\prime\right\rangle.\label{greenfunc:2}
\end{multline}
The propagators for the 15 unitary transformations are given by
\begin{eqnarray}
\left\langle x,y\left\vert\hat{U}_1^\dag\left(t\right)\right\vert x_1,y_1 \right\rangle
&=&\exp\left(-\frac{i}{\hbar}\alpha_1\right)\delta\left(x-x_1\right)\delta\left(y-y_1\right),\label{gen:prop1}\\
\left\langle x_1,y_1\left\vert\hat{U}_2^\dag\left(t\right)\right\vert x_2,y_2 \right\rangle
&=&\exp\left(-\frac{i}{\hbar}\alpha_2 x_2\right)\delta\left(x_1-x_2\right)\delta\left(y_1-y_2\right),\\
\left\langle x_2,y_2\left\vert\hat{U}_3^\dag\left(t\right)\right\vert x_3,y_3 \right\rangle
&=&\exp\left(-\frac{i}{\hbar}\alpha_3 y_3\right)\delta\left(x_2-x_3\right)\delta\left(y_2-y_3\right),\\
\left\langle x_3,y_3\left\vert\hat{U}_4^\dag\left(t\right)\right\vert x_4,y_4 \right\rangle
&=&\delta\left(x_3-x_4-\alpha_4\right)\delta\left(y_3-y_4\right),\\
\left\langle x_4,y_4\left\vert\hat{U}_5^\dag\left(t\right)\right\vert x_5,y_5 \right\rangle
&=&\delta\left(x_4-x_5\right)\delta\left(y_4-y_5-\alpha_5\right),\\
\left\langle x_5,y_5\left\vert\hat{U}_6^\dag\left(t\right)\right\vert x_6,y_6 \right\rangle
&=&\exp\left(-\frac{i}{\hbar}\alpha_6 x_6^2\right)\delta\left(x_5-x_6\right)\delta\left(y_5-y_6\right),\\
\left\langle x_6,y_6\left\vert\hat{U}_7^\dag\left(t\right)\right\vert x_7,y_7 \right\rangle
&=&\exp\left(-\frac{i}{\hbar}\alpha_7 y_7^2\right)\delta\left(x_6-x_7\right)\delta\left(y_6-y_7\right),\\
\left\langle x_7,y_7\left\vert\hat{U}_8^\dag\left(t\right)\right\vert x_8,y_8 \right\rangle
&=&\exp\left(-\frac{i}{\hbar}\alpha_8 x_8 y_8\right)\delta\left(x_7-x_8\right)\delta\left(y_7-y_8\right),\\
\left\langle x_8,y_8\left\vert\hat{U}_9^\dag\left(t\right)\right\vert x_9,y_9 \right\rangle
&=&\frac{\exp\left[\frac{i}{4\hbar\alpha_9}\left(x_8-x_9\right)^2\right]}{\sqrt{4\pi\hbar\alpha_9}}
\delta\left(y_8-y_9\right) , \\
\left\langle x_9,y_9\left\vert\hat{U}_{10}^\dag\left(t\right)\right\vert x_{10},y_{10} \right\rangle
&=&\delta\left(x_9-x_{10}\right) 
\frac{\exp\left[\frac{i}{4\hbar\alpha_{10}}\left(y_9-y_{10}\right)^2\right]}{\sqrt{4\pi\hbar\alpha_{10}}},\\
\left\langle x_{10},y_{10}\left\vert\hat{U}_{11}^\dag\left(t\right)\right\vert x_{11},y_{11}\right\rangle
&=&\frac{\exp\left\{\frac{i}{\hbar\alpha_{11}}\left[\left(x_{10}-x_{11}\right)y_{10}-y_{11}x_{10}\right]
\right\}}{2\pi\hbar\alpha_{11}},\\
\left\langle x_{11},y_{11}\left\vert\hat{U}_{12}^\dag\left(t\right)\right\vert x_{12},y_{12}\right\rangle
&=&\exp\left(-\alpha_{12}\right)\delta\left({\rm e}^{-2\alpha_{12}} 
x_{11}-x_{12}\right)\delta\left(y_{11}-y_{12}\right),
\\
\left\langle x_{12},y_{12}\left\vert\hat{U}_{13}^\dag\left(t\right)\right\vert x_{13},y_{13}\right\rangle
&=&\exp\left(-\alpha_{13}\right)\delta\left(x_{12}-x_{13}\right)\delta\left({\rm 
e}^{-2\alpha_{13}}y_{12}-y_{13}\right),
\\
\left\langle x_{13},y_{13}\left\vert\hat{U}_{14}^\dag\left(t\right)\right\vert x_{14},y_{14} \right\rangle
&=&\delta\left(x_{13}-x_{14}\right)\delta\left(y_{13}-y_{14}-\alpha_{14}x_{14}\right),\\
\left\langle x_{14},y_{14}\left\vert\hat{U}_{15}^\dag\left(t\right)\right\vert x^\prime,y^\prime 
\right\rangle
&=&\delta\left(x_{14}-x^\prime-\alpha_{15} y^\prime\right)\delta\left(y_{14}-y^\prime\right).\label{gen:prop15}
\end{eqnarray}
After substituting (\ref{gen:prop1})-(\ref{gen:prop15}) in to
the general expression for the propagator (\ref{greenfunc:2}) and
integrating, the Green function takes the final form
\begin{multline}
G(x,y,t; x^\prime, 
y^\prime,0)=\frac{\left(1+i\right)^{2}\eta}{4\pi\hbar\alpha_{11}}\exp\left(\alpha_{12}+\alpha_{13}
\right)\\
\times\exp\left\{-\frac{i}{\hbar}\left[\left(\frac{4\alpha_9\alpha_6-1}{4\alpha_9}\right)
\left(x-\alpha_4\right)^2+\alpha_7\left(y-\alpha_5\right)^2\right.\right.\\
\left.\left.+\alpha_8\left(x-\alpha_4\right)
\left(y-\alpha_5\right)+\frac{f}{\alpha_{11}}\left(y-\alpha_5\right)+\alpha_3 
y+\alpha_2 x+\frac{\alpha_{10}}{\alpha_{11}^2}f+\alpha_1\right]\right\}\\
\times\exp\left\{-\frac{i}{\hbar}\alpha_9\eta^{2}\left[\frac{x-\alpha_4}{2\alpha_9}
-\frac{y-\alpha_5}{\alpha_{11}}+\frac{1}{\alpha_{11}}\left(g-2\frac{\alpha_{10}}{\alpha_{11}}f\right)\right]
^2\right\},
\end{multline}
where
\begin{eqnarray}
f&=&\exp\left(2\alpha_{12}\right)\left(x^\prime+\alpha_{15}y^\prime\right),\\
g&=&\exp\left(2\alpha_{13}\right)\left(y^\prime+\alpha_{14}x^\prime+\alpha_{14}\alpha_{15}y^\prime\right),\\
\eta^{2}&=&\frac{\alpha_{11}^2}{\alpha_{11}^2-4\alpha_9 \alpha_{10}}.
\end{eqnarray}

\section{Two-dimensional charged particle subject to an in-plane electric field and
a perpendicular magnetic field}\label{twodimcharge}
In order to illustrate the use of the Lie algebraic approach let us study the dynamics
of a two-dimensional charged particle subject to an in-plane time dependent electric field and
a perpendicular magnetic field given by the Hamiltonian in Eq. (\ref{gfho}).
In this case $a_2=e E_x$, $a_3=e E_y$, $a_6=a_7=m\omega_c^2/8$,
$a_9=a_{10}=1/2m$, $a_{14}=-a_{15}=\omega_c/2$ and $a_1=a_4=a_5=a_8=a_{11}=a_{12}=a_{13}=0$
where  $\omega_c=eB/m$ is the cyclotron frequency.
For the sake of simplicity we consider the case where
$E_x$, $E_y$ and $B$ are constant although the more general
case where these quantities are time-dependent
can, in principle, be dealt with\cite{IbarraSierra201583}.
Substituting the previous parameters into the system of
ordinary differential equations
given by (\ref{difeq:01})-(\ref{difeq:15}) we obtain the
explicit form of the $\alpha$ parameters.
As stated above, the generators corresponding to the
first five parameters form a closed sub-algegbra of
$\mathcal L_{15}$ therefore the first five differential equations
(\ref{difeq:01})-(\ref{difeq:05}) may be solved independently
from the rest of the system. The solution to the first five
differential equations is
\begin{eqnarray}
\alpha_1 &=& \frac{e^2}{2m\omega_c^3}\left(E_x^2+E_y^2\right)
  \left(\sin\omega_ct-\omega_ct\cos\omega_ct\right),\nonumber \\
\alpha_2 &=& -\frac{eE_y}{2\omega_c}+\frac{e}{2}E_xt
+\frac{e}{2\omega_c}\left(E_x\sin\omega_ct+E_y\cos\omega_ct\right),\nonumber\\
\alpha_3 &=&\frac{eE_x}{2\omega_c}+\frac{e}{2}E_yt
+\frac{e}{2\omega_c}\left(E_y\sin\omega_ct-E_x\cos\omega_ct\right),\\
\alpha_4 &=& -\frac{eE_x}{m\omega_c^2}+\frac{e}{m\omega_c}E_yt
+\frac{e}{m\omega_c^2}\left(E_x\cos\omega_ct-E_y\sin\omega_ct\right),\nonumber\\
\alpha_5 &=& -\frac{eE_y}{m\omega_c^2}-\frac{e}{m\omega_c}E_xt
+\frac{e}{m\omega_c^2}\left(E_x\sin\omega_ct+E_y\cos\omega_ct\right),\nonumber
\end{eqnarray} 
where it can be easily verified that these functions correspond to the
classical solution for the position and momentum of a charged particle moving
in constant and uniform electromagnetic fields.
The generators $\hat h_6$, $\hat h_7$ and $\hat h_8$ also form a
sub-algebra of $\mathcal L_{15}$ and therefore yield three
differential equations that may be solved for $\alpha_6$, $\alpha_7$, $\alpha_8$,
apart from the remaining differential equations. By recasting Eqs. (\ref{difeq:06})-(\ref{difeq:08})
in terms of $\alpha_6-\alpha_7$, $\alpha_6+\alpha_7$ and $\alpha_8$ and reminding
that the initial conditions are $\alpha_6(0)=\alpha_7(0)=\alpha_8(0)=0$ the solution for these
three parameters is readily obtained as
\begin{align}
\alpha_6=\alpha_7=\frac{m\omega_c}{4}\tan\frac{\omega_ct}{2},&&
\alpha_8=0.\label{alpha:678}
\end{align}
The following three generators $\hat h_9$, $\hat h_{10}$ and $\hat h_{11}$ do not form
a closed algebra and therefore
the corresponding differential equations have
dependencies in parameters others than  $\alpha_9$, $\alpha_{10}$ and $\alpha_{11}$.
However, having obtained $\alpha_6$, $\alpha_7$ and $\alpha_8$ and by rewriting
Eqs. (\ref{difeq:09})-(\ref{difeq:11}) in terms of $\alpha_9-\alpha_{10}$, $\alpha_9+\alpha_{10}$
and $\alpha_{11}$ we find
\begin{align}
\alpha_9=\alpha_{10}=\frac{1}{m\omega_c}\cos\frac{\omega_ct}{2}\sin\frac{\omega_ct}{2},&&
\alpha_{11}=0,\label{alpha:91011}
\end{align}
where we have used  $\alpha_9(0)=\alpha_{10}(0)=\alpha_{11}(0)=0$.
After substituting the results for $\alpha_6$, $\alpha_7$, $\alpha_8$ in the differential equations
(\ref{difeq:12})-(\ref{difeq:14}) and rewriting for $\alpha_{13}-\alpha_{12}$,
$\alpha_{13}+\alpha_{12}$ and $\alpha_{14}$ the three differential equations yield the
following Riccati differential equation
\begin{equation}
\dot \beta-\beta^2-\left(\frac{\omega_c}{2}\right)^2=0,
\end{equation}
where $\beta=\dot\alpha_{13}-\dot\alpha_{12}$. Solving this equation and using the
initial conditions  $\alpha_{12}(0)=\alpha_{13}(0)=\alpha_{14}(0)=0$ the solution for
$\alpha_{12}$, $\alpha_{13}$ and $\alpha_{14}$
is readily found 
\begin{align}
\alpha_{12}=\ln\left(\cos\frac{\omega_ct}{2}\right),&&
\alpha_{13}=0,&&
\alpha_{14}=\cos\frac{\omega_ct}{2}\sin\frac{\omega_ct}{2}.
\label{alpha:121314}
\end{align}
The parameter $\alpha_{15}$ is calculated by direct integration of Eq. (\ref{difeq:15})
giving
\begin{align}
\alpha_{15}=-\tan\frac{\omega_ct}{2}.\label{alpha:15}
\end{align}
Upon replacing the explicit form of the parameters
$\alpha_6$-$\alpha_{15}$ in Eqs. (\ref{alpha:678}),
(\ref{alpha:91011}), (\ref{alpha:121314}) and (\ref{alpha:15})
in the general form for the
Heisenberg picture position and momentum operators 
(\ref{fho:xh:1})-(\ref{fho:pyh:1}) we get
\begin{eqnarray}
\hat x_H &=& \alpha_4+\cos\frac{\omega_ct}{2}
\left(\cos\frac{\omega_ct}{2}\hat x
  -\sin\frac{\omega_ct}{2}\hat y\right)\nonumber\\
&&\quad\quad\quad\quad\quad\quad\quad\quad
+\frac{2}{m\omega_c}\sin\frac{\omega_ct}{2}
\left(\cos\frac{\omega_ct}{2}\hat p_x
  -\sin\frac{\omega_ct}{2}\hat p_y\right),\\
\hat y_H &=& \alpha_5+\cos\frac{\omega_ct}{2}
\left(\sin\frac{\omega_ct}{2}\hat x
  +\cos\frac{\omega_ct}{2}\hat y\right)\nonumber\\
&&\quad\quad\quad\quad\quad\quad\quad\quad
+\frac{2}{m\omega_c}\sin\frac{\omega_ct}{2}
\left(\sin\frac{\omega_ct}{2}\hat p_x
  +\cos\frac{\omega_ct}{2}\hat p_y\right),\\
\hat p_{xH} &=& -\alpha_2
-\frac{m\omega_c}{2}\sin\frac{\omega_ct}{2}
\left(\cos\frac{\omega_ct}{2}\hat x
  -\sin\frac{\omega_ct}{2}\hat y\right)\nonumber\\
&&\quad\quad\quad\quad\quad\quad\quad\quad
+\cos\frac{\omega_ct}{2}
\left(\cos\frac{\omega_ct}{2}\hat p_x
  -\sin\frac{\omega_ct}{2}\hat p_y\right),\\
\hat p_{yH} &=& -\alpha_3
-\frac{m\omega_c}{2}
\sin\frac{\omega_ct}{2}
\left(\sin\frac{\omega_ct}{2}\hat x
  +\cos\frac{\omega_ct}{2}\hat y\right)\nonumber\\
&&\quad\quad\quad\quad\quad\quad\quad\quad
+\cos\frac{\omega_ct}{2}
+\left(\sin\frac{\omega_ct}{2}\hat p_x
  +\cos\frac{\omega_ct}{2}\hat p_y\right).
\end{eqnarray}
Finally, replacing the explicit form
of the $\alpha_6$-$\alpha_7$ parameters,
the general form of the propagator (\ref{greenfunc:2})
yields the expression 
\begin{multline}
G(x,y,t; x^\prime,y^\prime,0)=
\frac{m\omega_c}{4\pi\hbar \sin\frac{\omega_ct}{2}}
\exp\left[{-\frac{i}{\hbar}
\left(\alpha_1+\alpha_2 x+\alpha_3y\right)}\right]\\
\times\exp\Bigg[\frac{im\omega_c}{4\hbar\sin\frac{\omega_ct}{2}}
\bigg\{\Big[\left(x-\alpha_4\right)^2+\left(y-\alpha_5\right)^2
+(x^{\prime})^2+(y^\prime)^2\Big]\cos\frac{\omega_ct}{2}\\
-2\cos\frac{\omega_ct}{2}x^\prime\left(x-\alpha_4\right)
-2\sin\frac{\omega_ct}{2}x^\prime\left(y-\alpha_5\right)\\
+2\sin\frac{\omega_ct}{2}y^\prime\left(x-\alpha_4\right)
-2\cos\frac{\omega_ct}{2}y^\prime\left(y-\alpha_5\right)
\bigg\}\Bigg].
\end{multline}
Comparable results for the propagator and the Heisenberg picture
position and momentum operator have been obtained via
the path integral method  \cite{0305-4470-17-4-022}
or time-dependent perturbation approach of the
Fock-Darwin Hamiltonian \cite{PhysRevA.80.053401}.

\section{Conclusions} \label{conclusions}
We have developed a systematic method based on the Lie algebraic
approach to obtain the evolution operator and its corresponding 
propagator for the generalized two-dimensional quadratic Hamiltonian.
This method relies on the possibility of expressing the Hamiltonian
as a linear combination of elements that form Lie algebra with coefficients
that in general are time-dependent functions.
In this case the evolution operator is a member of the
unitary group generated by these elements, and therefore is expressible
in terms of the elements of the same algebra and the corresponding
time-dependent transformation parameters.

Finding the explicit time-dependence of the the transformation
parameters determines completely the evolution
operator. Therefore we have exploited the properties of
Hamiltonians having a dynamical algebra to find analytical expressions
for the ordinary differential equations that govern
the dynamics of the transformation parameters. 

Even though the method presented here is mainly intended
to obtain the evolution operator for the generalized
two-dimensional quadratic Hamiltonian, the results
from Section \ref{liealap} are general enough that may be applied to any
Hamiltonian having a dynamical algebra with a large
number of elements.

To illustrate the method we have presented
the example of a two-dimensional charged particle in
uniform electro-magnetic fields.
The obtained propagator and
Heisenberg picture position and momentum operators
are consistent with the ones calculated with the
path integral method  \cite{0305-4470-17-4-022}
and time-dependent perturbation theory of the
Fock-Darwin Hamiltonian \cite{PhysRevA.80.053401}.

The rather general form of the two-dimensional quadratic
Hamiltonian allows this method to
tackle a wide variety of significant physical situations
such as two-dimensional single electrons
trapped inasymmetric quantum dots with parabolic confinement, or
charged particle subject to time-varying uniform
electro-magnetic fields among others.

\section{Aknowledgments}  
The authors would like to thank the
``Departamento de Ciencias B硬icas UAM-A" for the
financial support.
J. C. Sandoval-Santana and V. Ibarra-Sierra
would like to acknowledge the support
received from ``Becas de Posgrado UAM".

\appendix

\section{Usefull relations}

Since it is widely used to calculate most of the transformation
rules, we enunciate the next commutation relation.
If the commutor 
\begin{equation}
 \left[\hat{A},\hat{B}\right]=\hat{C},
\end{equation}
commutes with the operators $\hat A$ and $\hat B$, i. e.
\begin{equation}
\left[\hat{A},\hat{C}\right]=\left[\hat{B},\hat{C}\right]=0.
\end{equation}
then it follows that
\begin{equation}\label{re:cuan}
\left[\hat{A},F(\hat{B})\right]=\left[\hat{A},\hat{B}\right]\frac{\partial
F(\hat{B})}{\partial\hat{B}},
\end{equation}
provided that $F$ is an analytical function.


%

\end{document}